\documentclass[11pt]{article}
\usepackage{epsf,amsmath,amssymb,axodraw,graphicx,scalefnt}
\usepackage{color}
\usepackage{rotating}
\newcommand{\qcd}{{\abbrev QCD}}
\newcommand{\code}{\tt}

\newcommand{\abbrev}{\scalefont{.9}\rm}
\newcommand{\lnx}[1]{\ln^{#1} x}
\newcommand{\lnxm}[1]{L^{#1}(x)}

\newcommand{\lht}{l_{Ht}}
\newcommand{\muF}{\mu_{\rm F}}
\newcommand{\muR}{\mu_{\rm R}}
\newcommand{\mhiggs}{M_{\rm H}}
\newcommand{\mtop}{M_{t}}
\newcommand{\ep}{\epsilon}
\newcommand{\api}{\frac{\alpha_s}{\pi}}
\newcommand{\eqn}[1]{Eq.\,(\ref{#1})}
\newcommand{\fig}[1]{Fig.\,\ref{#1}}

\newcommand{\dd}{{\rm d}}

\newcommand{\order}[1]{{\cal O}(#1)}
\newcommand{\lhc}{{\abbrev LHC}}
\newcommand{\sm}{{\abbrev SM}}

\newcommand{\lo}{{\abbrev LO}}
\newcommand{\nlo}{{\abbrev NLO}}
\newcommand{\nnlo}{{\abbrev NNLO}}

\newcommand{\msbar}{\overline{\mbox{\abbrev MS}}}

\newcommand{\bld}[1]{\boldmath{$#1$}}

\def\readRCS$#1: #2,v #3 #4 #5${%
\def\filename{#2}%
\def\fileversion{#3}%
\def\filedate{#4}%
}

\readRCS
$Id: gghmt.tex,v 1.1 2009/04/17 08:46:42 rharland Exp $

\date{}
\title{\vspace*{-6em}
  \begin{flushright}
    {\sf\small September 2009 --- WUB/09-10}
  \end{flushright}
  \vspace*{2em}Finite top mass effects for hadronic Higgs production at next-to-next-to-leading order} \author{Robert V. Harlander and Kemal J.
  Ozeren\\[2em] {\it Fachbereich C, Bergische Universit\"at
    Wuppertal}\\{\it 42097 Wuppertal, Germany}\\ {\tt
    robert.harlander@uni-wuppertal.de}\\ {\tt
    ozeren@physik.uni-wuppertal.de} }

\begin{document}
\maketitle

\begin{abstract}
The first four terms of an expansion in $\mhiggs^2/\mtop^2$ of the total
inclusive cross section for Higgs production in gluon fusion are
evaluated through next-to-next-to-leading order \qcd{}. A reliable and
precise approximation of the full top mass dependence at \nnlo{} is
derived and compared to the frequently used heavy-top limit. It is found
that both results agree numerically to better than 0.5\% in the Higgs mass
range of 100-300~GeV. This validates the higher order results for the
inclusive Higgs cross section and justifies the heavy-top limit as a
powerful tool for Higgs phenomenology at the \lhc{} and the Tevatron.
\end{abstract}

\section{Introduction}

It is expected that the Large Hadron Collider (\lhc{}) will provide
insight into the mechanism of electro-weak symmetry breaking. Many of
the theoretical models, including the Standard Model (\sm{}), implement
it by introducing an as of yet unobserved elementary scalar particle, a
so-called Higgs boson. One of the tasks of the \lhc{} is therefore to
search for a Higgs boson and to measure its properties (for reviews, see
Refs.~\cite{hunter,Djouadi:2005gi,Djouadi:2005gj}).

Precise measurements will require a thorough understanding of the Higgs
production cross section. In this respect, the gluon fusion process
poses a number of problems: the gluon parton densities need to be known
very precisely; the cross section is of order $\alpha_s^2$ and thus very
sensitive to the precise value of this quantity; the radiative
corrections and thus the estimated uncertainty from higher order effects
is unusually large.

The latter aspect is particularly problematic because the gluon fusion
process is loop induced, so that the next-to-leading order (\nlo{})
effects already require a two-loop calculation. Fortunately, it was
found that the \nlo{} $K$-factor is extremely well reproduced in an
effective theory approach, obtained by integrating out the top quark
from the theory. Details will be discussed below.

Based on this observation, it is commonly assumed that the \nnlo{}
corrections, obtained within this effective theory approach, are
practically equivalent to the full calculation, at least below the top
quark threshold. In fact, almost all radiative corrections beyond \nlo{}
have been treated in this framework up to now.

In this paper, we study the quality of this approximation by evaluating
subleading terms in the expansion in $1/\mtop$. After introducing our
notation and the basic formulas in Section~\ref{sec::gluonfusion}, we
describe our calculational methods in Section~\ref{sec::mtterms}, while
the analytic results for the $1/\mtop{}$ expansion are presented in
Section~\ref{sec::anares}. A matching of the low- to the high energy
region is described in Section~\ref{sec::smallx} and the numerical
analysis is presented in Section~\ref{sec::numerics}. The conclusions of
our findings are drawn in Section~\ref{sec::conclusions}.

\section{Higgs production through gluon fusion}\label{sec::gluonfusion}

The inclusive hadronic cross section for \sm{} Higgs production in
proton--(anti-)proton collisions is obtained by convoluting the partonic
cross section $\hat\sigma_{\alpha\beta\to H+X}$ for the scattering of
parton $\alpha$ with parton $\beta$ by the corresponding parton density
functions $\phi_{\alpha/p}(x)$ ({\abbrev PDF}s):
\begin{equation}
\begin{split}
\sigma_{pp'\to H+X}(s) &= \sum_{\alpha,\beta\in\{q,\bar q,g\}}\int_0^1\dd \tau \,
    {\cal E}_{\alpha\beta}(\tau)\,\hat\sigma_{\alpha\beta\to
  H+X}(\hat s=\tau s)\,,\\
{\cal E}_{\alpha\beta}(\tau) &\equiv \int_\tau^1\frac{\dd
  x}{x}\left[ \phi_{\alpha/p}(x)\phi_{\beta/p'}(\tau/x)\right]\,,\qquad
p'\in \{p,\bar p\}\,.
\label{eq::sigmapp}
\end{split}
\end{equation}
(In the following, we will shorten or drop the subscripts on ${\cal E}$,
$\phi$, $\sigma$, and $\hat\sigma$ whenever there is no chance of
confusion).  In this formula, we have suppressed the dependence on the
factorization scale $\muF$ for $\phi$ and $\hat \sigma$, as well as the
dependence of $\sigma$ and $\hat\sigma$ on $\mtop$ and $\mhiggs$. The
partonic cross section $\hat\sigma$ is evaluated in terms of a
perturbation series in $\alpha_s$. Its leading order contribution arises
from the triangle diagram shown in \fig{fig::dias}\,(a) (plus the one
with opposite fermion direction), where the fermion line can in
principle be any of the six quarks. However, due to the Yukawa
couplings, the by far dominant contribution is due to top quarks, while
the bottom quark has an effect of rougly $7\%$ at the \lhc{}\footnote{By
  ``\lhc{}'', we refer to $pp$ collisions at 14\,TeV center-of-mass
  energy, although the initial energy of the \lhc{} will be lower; more
  detailed phenomenological studies at various energies are deferred to
  a future publication.}, and $9\%$ at the Tevatron. All lighter quarks
can safely be neglected.

\nlo{} \qcd{} corrections to the inclusive cross section have been known
for a long time. They were first evaluated for the top quark induced
terms in the so-called ``heavy-top limit'' which will be described in
more detail below, and were found to increase the cross section by
roughly 70\% at the \lhc{} w.r.t.\ the leading order
prediction~\cite{Dawson:1990zj,Djouadi:1991tk}. This was confirmed by
the general result for arbitrary top and bottom quark
mass~\cite{Graudenz:1992pv,Spira:1995rr}.

The large \nlo{} effects clearly asked for the evaluation of the \nnlo{}
corrections which, considering its success at \nlo{}, are generally
assumed to be well approximated in the heavy-top limit. They lead to another
significant increase of the total cross
section~\cite{Harlander:2002wh,Anastasiou:2002yz,Ravindran:2003um}, so
that its actual value turns out to be roughly a factor of two above the
\lo{} prediction for the \lhc{}, and even up to a factor of three at the
Tevatron (for the latest compilations, see
Refs.\,\cite{deFlorian:2009hc,Anastasiou:2008tj}). Further studies that
go beyond the fixed-order \nnlo{} result have not found significant
effects and have thus corroborated the stability of the perturbative
series (see, e.g.~Refs.\,\cite{Catani:2003zt,Ahrens:2008qu}).

The goal of this paper is to go beyond the heavy-top limit in order to
test the quality of this approximation. In principle, our approach would
also allow us to derive an improved prediction of the inclusive cross
section. However, it will turn out that the heavy-top limit works so
well that it reproduces our improved result to better than 0.5\%
accuracy.

We write the top quark induced partonic cross section as
\begin{equation}
\begin{split}
\hat\sigma_{\alpha\beta} &= \sigma_0\,\Delta_{\alpha\beta}\,,
\label{eq::sigmahat}
\end{split}
\end{equation}
with
\begin{equation}
\begin{split}
\sigma_0 &= \frac{\pi\sqrt{2}G_{\rm F}}{256}\left(\api\right)^2
\tau^2\left|1+(1-\tau)\arcsin^2\frac{1}{\sqrt{\tau}}\right|^2\,,\qquad
\tau = \frac{4\mtop^2}{\mhiggs^2}\,.
\label{eq::sigma0}
\end{split}
\end{equation}
Here, $G_{\rm F}\approx 1.16637\cdot 10^{-5}$\,GeV$^{-1}$ is Fermi's
constant, and throughout this paper, $\mtop{}$ denotes the on-shell top
quark mass, and $\alpha_s\equiv \alpha_s^{(5)}(\muR)$ the strong
coupling for five active quark flavours at the renormalization scale
$\muR$. The kinetic terms assume the form
\begin{equation}
\begin{split}
\Delta_{\alpha\beta}(x) = \delta_{\alpha g}\delta_{\beta g}\,\delta(1-x) +
\api\Delta^{(1)}_{\alpha\beta}(x) + 
\left(\api\right)^2\Delta^{(2)}_{\alpha\beta}(x) + \ldots\,,
\end{split}
\end{equation}
where
\begin{equation}
\begin{split}
x = \mhiggs^2/\hat s\,.
\end{split}
\end{equation}
The $\Delta^{(n)}$ still depend on $\mtop$, and logarithmically on the
renormalization and factorization scales $\muR$ and $\muF$. At \nlo{},
the full $\mtop$ dependence is known in numerical
form~\cite{Graudenz:1992pv} (the virtual terms are known
analytically~\cite{Harlander:2005rq,Anastasiou:2006hc,Aglietti:2006tp}). At
\nnlo{}, only the heavy-top approximation is known,
however~\cite{Harlander:2002wh,Anastasiou:2002yz,Ravindran:2003um}. It
will be discussed in more detail in the next section.

Note that in order to arrive at a consistent \nnlo{} result for the
hadronic cross section, one needs to evaluate \eqn{eq::sigmapp} not only
with an \nnlo{} expression for $\hat\sigma$, but also by taking into
account expressions for $\alpha_s$ and the {\abbrev PDF}s at the
appropriate order. In this paper, we use the central set of {\abbrev
  MSTW2008}~\cite{Martin:2009iq} which are the latest available \nnlo{}
{\abbrev PDF}s. A detailed study of the {\abbrev PDF} uncertainties is
left for a future publication.

Apart from these pure \qcd{} corrections, also the leading electro-weak
effects have been
evaluated~\cite{Djouadi:1994ge,Aglietti:2004nj,Degrassi:2004mx,Actis:2008ug},
and an estimate of the mixed electro-weak/\qcd{} corrections is
available~\cite{Anastasiou:2008tj} as well.

\section{Calculation of the \bld{1/\mtop{}} terms}\label{sec::mtterms}

It is well known that the gluon-Higgs interaction in the heavy-top limit
can be expressed in terms of an effective Lagrangian~\cite{Ellis:1975ap}:
\begin{equation}
\begin{split}
{\cal L}_{\rm eff} &= -\frac{H}{4v}\,C_1\,
G_{\mu\nu}^aG^{\mu\nu,a} + {\cal L}_\qcd^{(5)}\,,
\label{eq::leff}
\end{split}
\end{equation}
where $v=246$\,GeV and the Wilson coefficient $C_1$ is meanwhile known
through
$\order{\alpha_s^{5}}$~\cite{Chetyrkin:1997iv,Schroder:2005hy,Chetyrkin:2005ia}. For
completeness, we quote it here through
$\order{\alpha_s^{3}}$~\cite{Chetyrkin:1997un,Kramer:1996iq} which is
sufficient at \nnlo{}:
\begin{equation}
\begin{split}
C_1 =& -\frac{1}{3}\api\bigg\{
  1+\frac{11}{4}\api + \left(\api\right)^2\bigg[
  \frac{2777}{288} + \frac{19}{16}\ln\frac{\mu^2}{\mtop^2} +
  n_l\bigg(-\frac{67}{96} +
  \frac{1}{3}\ln\frac{\mu^2}{\mtop^2}\bigg)\bigg]\bigg\}\,.
\label{eq::coefc}
\end{split}
\end{equation}
${\cal L}_\qcd^{(5)}$ is the \qcd{} Lagrangian with $n_l=5$ massless quark
flavours.

Instead of working strictly within this effective theory, however, one
usually factors out the full leading order top mass dependence from the
Higgs production cross section and writes
\begin{equation}
\begin{split}
\sigma_{\infty}(s) &= \sum_{\alpha,\beta}\int_0^1\dd \tau \,
    {\cal E}_{\alpha\beta}(\tau)\,\hat\sigma_{\alpha\beta,\infty}(\tau s)\,,\qquad
\hat\sigma_{\alpha\beta,\infty} \equiv \sigma_0\,\Delta_{\alpha\beta,\infty}\,,
\label{eq::heavytop}
\end{split}
\end{equation}
where $\sigma_0$ is given in \eqn{eq::sigma0}, and
$\Delta_{\alpha\beta,\infty}$ is evaluated on the basis of
\eqn{eq::leff}. It is then clear that $\Delta^{(1)}_\infty$ does not
depend on $\mtop{}$, while $\Delta^{(2)}_\infty$ only has a logarithmic
$\mtop$-dependence through $C_1(\alpha_s)$.  Note that with this {\it
  definition of the heavy-top limit}, where the full $\mtop$-dependence
in $\sigma_0$ is kept, the \lo{} cross section is identical to the one
in the full theory.

It was observed long ago~\cite{Kramer:1996iq} that the total cross
section at \nlo{} in \qcd{} is approximated by $\sigma^\nlo_\infty$ to
better than 1\% up to values of $\mhiggs=2\mtop$; even at
$\mhiggs=1$\,TeV, the deviation to the exact result remains below 10\%
(see, e.g., Ref.\,\cite{Harlander:2003xy}).  This precision is quite
remarkable, because at higher orders, gluon fusion is a 3-scale process,
depending on $\mtop$, $\mhiggs$, and the center-of-mass (c.m.) energy
$\sqrt{\hat s}$. While at \lo{} $\hat s$ is fixed to $\mhiggs^2$, the
real radiation of quarks and gluons starting at \nlo{} allows
$\sqrt{\hat s}$ to vary from $\mhiggs$ up to the hadronic c.m.~energy
which may reach 14\,TeV at the \lhc{}. Considering more exclusive
quantities such as $p_T$-distributions (see Ref.\,\cite{Balazs:2004rd}
and references therein), or even fully differential approaches as in
Refs.\,\cite{Anastasiou:2004xq,Catani:2008me}, the number of scales
increases further and the heavy-top approximation may no longer be under
control. Recent studies in this direction can be found in
Refs.~\cite{Keung:2009bs,Anastasiou:2009kn}.

For the inclusive cross section -- which is the subject of this paper --
it is usually argued that the reason for the high quality of the
approximation in \eqn{eq::heavytop} is the dominance of soft gluon
radiation in the higher order corrections, but a solid theoretical
justification and a quantitative error estimate are still unavailable.
One may therefore remain in doubt about the applicability of
\eqn{eq::heavytop} beyond \nlo{}.  A way to test it is to evaluate
subleading terms in $1/\mtop{}$ to the partonic cross section, similar
to what was done at \nlo{} in Ref.\,\cite{Dawson:1993qf}. The effective
Lagrangian of \eqn{eq::leff} could clearly be extended to incorporate
such terms by including higher dimensional operators (see, e.g.,
Ref.\,\cite{Neill:2009tn}). However, in general at each order in
$1/\mtop{}$ the number of operators grows, and renormalization becomes
more and more clumsy, for example.

In this paper, instead of constructing an effective Lagrangian, we
directly evaluate the Feynman diagrams obtained from the six-flavour
Lagrangian by applying the method of asymptotic expansions (for a
review, see Ref.\,\cite{Smirnov:2002pj}). At \nnlo{}, the contributing
diagrams are at the 3-loop level for the process $gg\to H$, at the
2-loop level for the single real emission processes $gg\to Hg$, $qg\to
Hq$, $q\bar q\to Hg$, and at the 1-loop level for the double real
emissions $gg\to Hgg$, $gg\to Hq\bar q$, $qg\to Hqg$, $q\bar q\to Hq\bar
q$, $qq\to Hqq$ (identical quark flavours), and $qq'\to Hqq'$ (different
quark flavours; it is understood that the charge conjugated processes
need to be taken into account as well).  Each Feynman diagram
considered here contains at least one top quark loop. Examples for the
various contributions are shown in \fig{fig::dias}.

\newlength{\diasize}
\setlength{\diasize}{.22\textwidth}
\begin{figure}
  \begin{center}
    \begin{tabular}{ccc}
      \raisebox{.2em}{\includegraphics[width=1.1\diasize]{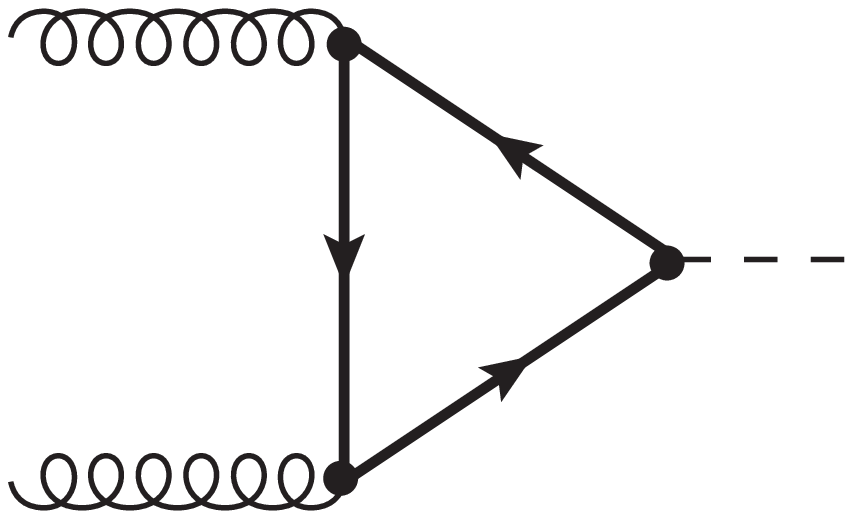}} &
      \includegraphics[width=1.1\diasize]{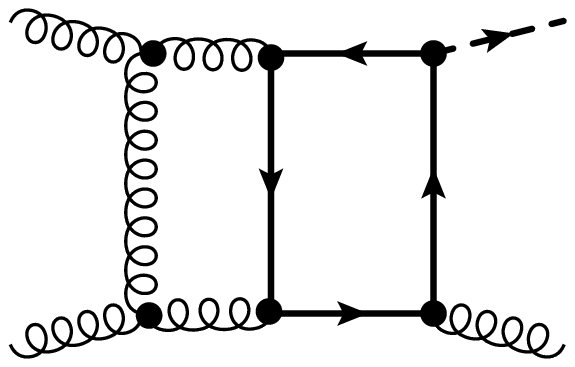} &
      \raisebox{.2em}{\includegraphics[width=\diasize]{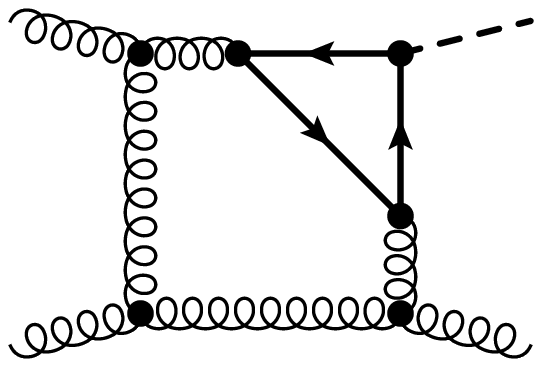}}\\
      (a) & (b) & (c)      \\[1em]
    \end{tabular}
    \begin{tabular}{cc}
      \raisebox{.2em}{\includegraphics[width=\diasize]{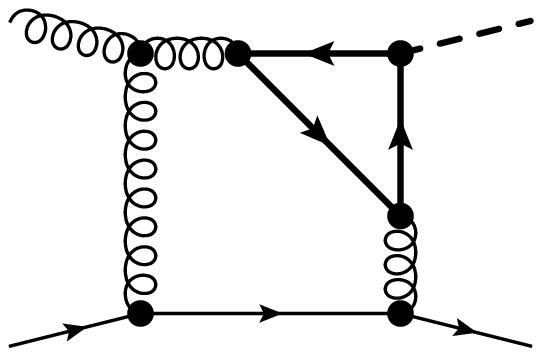}} &
      \includegraphics[width=1.1\diasize]{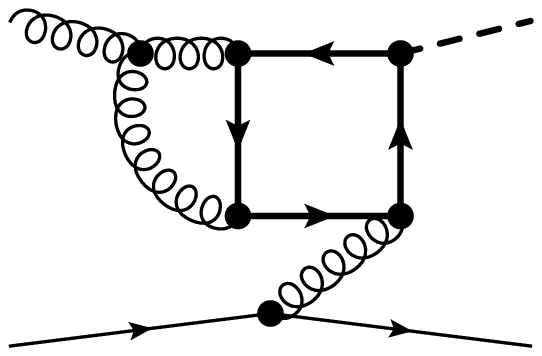}\\
      (d) & (e)\\[1em]
    \end{tabular}
    \begin{tabular}{ccc}
      \raisebox{0em}{\includegraphics[width=\diasize]{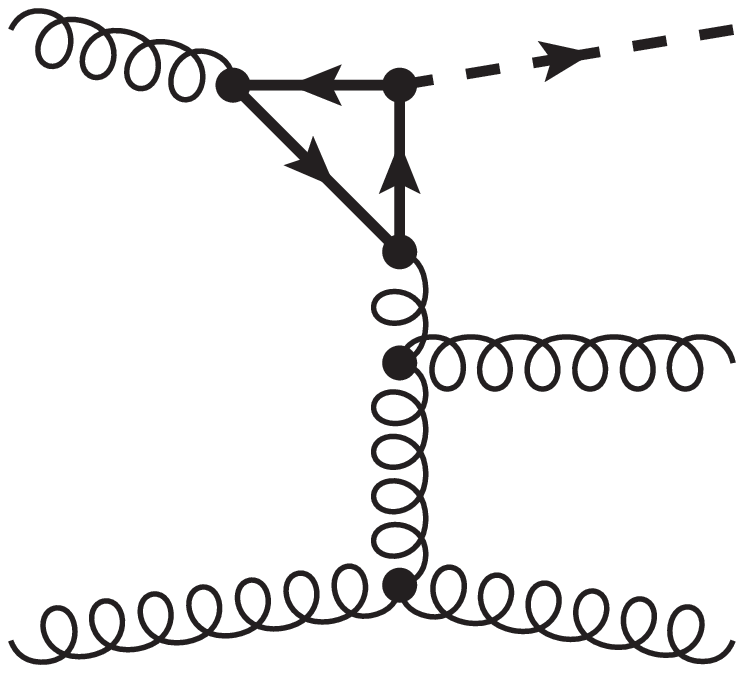}} &
      \raisebox{0em}{\includegraphics[width=\diasize]{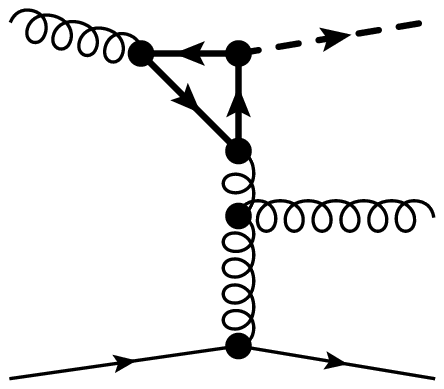}} &
      \includegraphics[width=\diasize]{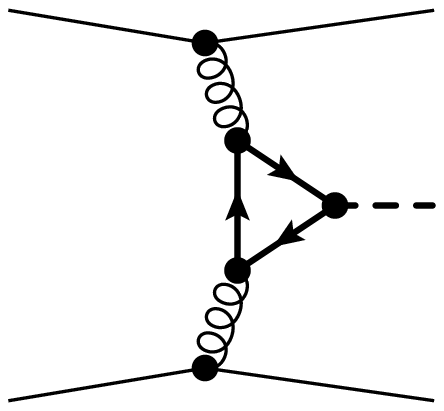}\\
       (f) & (g) & (h)
    \end{tabular}
    \parbox{.9\textwidth}{
      \caption[]{\label{fig::dias}\sloppy (a) \lo{} Feynman diagram for
        the gluon fusion process; (b)-(h) Sample Feynman diagrams
        contributing to the inclusive \nnlo{} cross section for Higgs
        production in gluon fusion. (b)-(e) single real radiation;
        (f)-(h) double real radiation.
    }}
  \end{center}
\end{figure}

In our approach, we assume $\mtop$ heavier than any other basic scale in
the process. This allows us to express all Feynman integrals as
convolutions of massive vacuum integrals with at most three loops, and
massless 3/4/5-point functions through 2/1/0 loops, respectively.  In
fact, the {\it purely virtual contributions} at \nnlo{} have already
been calculated using this
method~\cite{Harlander:2009bw,Pak:2009}\footnote{Compared to
  Ref.\,\cite{Harlander:2009bw}, we have additionally calculated the
  $1/\mtop^6$ term and found agreement with Ref.\,~\cite{Pak:2009}.}, so
we will not discuss them in more detail at this point.

For the {\it double real emission} contributions, the asymptotic
expansion is equivalent to the interchange of loop integration and
Taylor expansion in $p/\mtop$, where $p$ is any component of the
external momenta, and thus one is left only with 1-loop massive tadpole
integrals which can be easily evaluated with the help of {\tt
  MATAD}~\cite{Steinhauser:2000ry}. The difficulty arises from the phase
space integration which we perform in terms of an expansion around
$x=1$~\cite{Harlander:2002wh} (``soft expansion''). This is fully
justified due to the fact that the $1/\mtop$-expansion assumes
$\sqrt{\hat s}<2\mtop$ and thus $x\gtrsim 0.1$ anyway. Apart form that,
for the $1/\mtop^0$-terms, it was observed that the hadronic cross
section converges very well to the full result when successively higher
order terms in $(1-x)$ are included in the partonic cross
section~\cite{Harlander:2002wh}.

\begin{figure}
  \begin{center}
    \begin{tabular}{c}
      \mbox{\includegraphics[width=\diasize]{figs/gghg2l2.eps}}\quad
      \raisebox{.3\diasize}{$\to$}\quad
      \raisebox{.04\diasize}{\includegraphics[width=.84\diasize]{%
          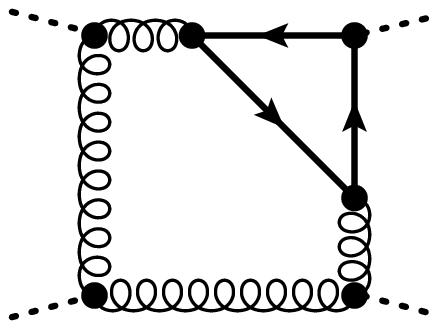}}\quad
      \raisebox{.3\diasize}{$\otimes$}
      \raisebox{.1\diasize}{%
        \includegraphics[width=.7\diasize]{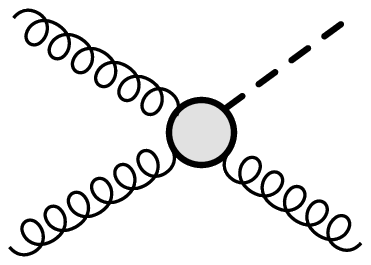}}\\[2em]
      \raisebox{.3\diasize}{$+$}
      \raisebox{.08\diasize}{%
        \includegraphics[width=.7\diasize]{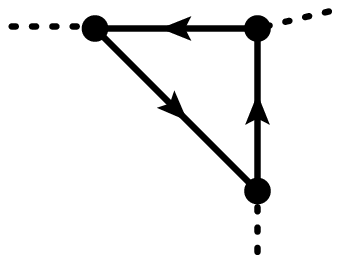}}\quad
      \raisebox{.3\diasize}{$\otimes$}
      \raisebox{.03\diasize}{%
        \includegraphics[width=.9\diasize]{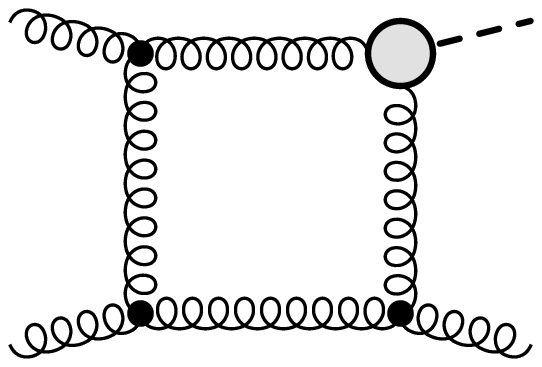}}
    \end{tabular}
    \parbox{.9\textwidth}{
      \caption[]{\label{fig::asympt1}\sloppy Diagrammatic representation
        of the asymptotic expansion of a particular Feynman diagram in
        the limit $\hat s, \mhiggs^2\ll 4\mtop^2$. The diagrams left of
        $\otimes$ represent subdiagrams of the original diagram that are
        to be expanded in the momenta corresponding to the dotted
        external lines before the loop integration. In this way, it is
        apparent that the original integral, depending on $\hat s$,
        $\mhiggs^2$ and $\mtop^2$, is decomposed into products of
        ``tadpole'' integrals with vanishing external momenta and
        massless four-point functions. The shaded blob in the diagrams right
        of $\otimes$ represents an effective vertex given by the result
        of the diagram left of $\otimes$ (for details of asymptotic
        expansions, see Ref.\cite{Smirnov:2002pj}, for example).}}
  \end{center}
\end{figure}

The graphical representation of the asymptotic expansion of one of the
{\it single real emission} diagrams is shown in \fig{fig::asympt1}. The
resulting integrals can be calculated by standard means: For the massive
tadpoles, we use {\code MATAD}~\cite{Steinhauser:2000ry}, and the 1-loop
tensor integrals are solved using Schwinger parametrisation and
integration-by-parts~\cite{Chetyrkin:1981qh}. Both routines are embedded
in the {\tt q2e/exp} framework~\cite{Q2E,Harlander:1997zb} which, in
combination with the diagram generator {\tt
  qgraf}~\cite{Nogueira:1993ex}, provides a fully automatic way to
calculate the relevant diagrams, including the procedure of asymptotic
expansions. This setup allows for the evaluation of arbitrary orders in
the $1/\mtop$ expansion, with the only limitation arising from the
available computing power. For this publication, we found that the
optimal cost/benefit ratio\footnote{As the measure of ``benefit'' we
  considered the numerical difference between the highest two terms in
  the $1/\mtop$ expansion.}  is reached at $\order{1/\mtop^6}$, i.e.,
three orders beyond the heavy-top limit known so far.  After
multiplication by the corresponding \nlo{} diagrams, we evaluate the
phase space integrals in terms of hypergeometric functions depending on
$x=\mhiggs^2/\hat s$ and $\ep=(4-D)/2$. Expansion around $\ep=0$ leads
to a Laurent series with poles at $\ep=0$ through $\order{1/\ep^4}$, and
coefficients depending on $x$ through polylogarithms up to the $4^{\rm
  th}$ degree.  Since we calculated the double real emission
contributions as expansions in $(1-x)$, we have to expand the single
real emission terms in the same way. This is of course trivial from the
full $x$-dependent result. However, also here we can perform the soft
expansion before the phase space integration which serves as a useful
check.

In the sum of the virtual, single-real, and double-real emission
contributions, the $1/\ep^4$ poles cancel, and upon {\abbrev
  UV}-renormalization of $\alpha_s$ (for which we adopt the $\msbar{}$
scheme), $\mtop{}$ and the gluon wave function (both in the on-shell
scheme), also the $1/\ep^3$-terms drop out. The remaining poles are of
infra-red nature, and are absorbed into the {\abbrev PDF}s with the help
of the usual mass factorization procedure. The relevant convolution
integrals are calculated by transforming them into Mellin space (which
turns them into simple products) and subsequent inverse Mellin
transformation with the help of Ref.\,\cite{Blumlein:1998if}.

\section{Analytical results}\label{sec::anares}

A natural decomposition of the kinetic terms to the inclusive cross
section is as follows:
\begin{equation}
\begin{split}
\Delta^{(n)}_{\alpha\beta} &= \delta_{\alpha g}\delta_{\beta g}\left[
  a^{(n)}\delta(1-x) + \sum_{k=0}^{2n-1}b_k^{(n)}{\cal
  D}_k(x)\right] + h_{\alpha\beta}^{(n)}(x)\,,
\label{eq::softhard}
\end{split}
\end{equation}
where the ${\cal D}_k(x)$ are the usual plus-distributions defined as
\begin{equation}
\begin{split}
\int_0^1\dd x\,{\cal D}_k(x)\,f(x) \equiv
\int_0^1\dd x\,\left[f(x) - f(1)\right]\frac{\ln^k(1-x)}{1-x}
\end{split}
\end{equation}
for arbitrary functions $f(x)$ (differentiable at $x=1$). The terms
arising from $a^{(n)}$ and $b_k^{(n)}$ are called the ``soft+virtual
contribution'', while $h_{\alpha\beta}^{(n)}$ is referred to as the
``hard contribution''. The $a^{(n)}$, $b_k^{(n)}$, and
$h_{\alpha\beta}^{(n)}(x)$ are functions of the top quark mass. We write
them as
\begin{equation}
\begin{split}
\{a,b_k,h_{\alpha\beta}(x)\} = \sum_{i\geq
  0}\left(\frac{\mhiggs}{\mtop}\right)^i
\{a_{i},b_{k,i},h_{\alpha\beta,i}(x)\}\,.
\label{eq::massexp}
\end{split}
\end{equation}
The leading terms in $1/\mtop$, i.e.\ $i=0$, have been obtained in
Refs.\,\cite{Harlander:2000mg,Harlander:2001is,Catani:2001ic}
(soft+virtual) and
Refs.\,\cite{Harlander:2002wh,Anastasiou:2002yz,Ravindran:2003um}
(hard).\footnote{In fact, the leading logarithmic hard term was obtained
  before in Ref.\,\cite{Kramer:1996iq}.}

Concerning the subleading terms in $1/\mtop$, let us begin with the
\nlo{} result (terms with odd powers of $\mhiggs/\mtop$ vanish):
\begin{equation}
\begin{split}
a^{(1)}_0 &= \frac{11}{2} + 6\zeta_2\,,\qquad
a^{(1)}_2 = \frac{34}{135}\,,\qquad
a^{(1)}_4 = \frac{3553}{113400}\,,\qquad
a^{(1)}_6 = \frac{917641}{190512000}\,,\\
b^{(1)}_{0,i} &= 0\quad\forall\ i\geq 0\,,\qquad
b^{(1)}_{1,0} = 12\,,\qquad
b^{(1)}_{1,i} = 0\quad\forall\ i\geq 2\,,
\label{eq::soft1}
\end{split}
\end{equation}
where $\zeta_2 \equiv \pi^2/6$, and we have set $\muF=\muR=\mhiggs$. For
completeness, we have also listed the leading terms in $1/\mtop$.  Note
that the coefficients of the plus distributions are fully determined in
the heavy-top limit, i.e., they do not receive $1/\mtop$
corrections~\cite{Graudenz:1992pv,Dawson:1993qf}. $b^{(1)}_{0,0}$
becomes non-zero for $\muF\neq \mhiggs$ as is obvious from the evolution
equations for $\muF$.

The hard terms read, for the $gg$ channel:
\begin{equation}
\begin{split}
h^{(1)}_{gg,0} &= -\frac{6\,(1 - x + x^2)^2}{1 - x}\lnx{}
-12x(2-x+x^2)\lnxm{} - \frac{11}{2}{(1-x)^3}
\,,\\ h^{(1)}_{gg,2} &= -\frac{3x(1-x)}{20}\,,\qquad
h^{(1)}_{gg,4} = \frac{(1-x)\left( 37 + 76\,x - 83\,x^2 +
68\,x^3 \right)}{11200\, x^2}\,,\\ 
h^{(1)}_{gg,6} &= \frac{(1 - x)\left(3864 - 4008\,x + 4133\,x^2 -
  2668\,x^3 + 3770\,x^4\right)}{2016000\,x^3}\,,
\label{eq::hgg1}
\end{split}
\end{equation}
where
\begin{equation}
\begin{split}
L(x) \equiv \ln(1-x)\,.
\end{split}
\end{equation}

For the $qg$ channel (which is identical to the $\bar qg$ channel), one
finds:
\begin{equation}
\begin{split}
h^{(1)}_{qg,0} &= \frac{2}{3}\,(2 - 2\,x + x^2)\,\left( 2\lnxm{} - \lnx{} \right)
-1 + 2\,x - \frac{x^2}{3}\,,\\
h^{(1)}_{qg,2} &= \frac{11(-4 + 6\,x - 3\,x^2 + x^3)}{270\,x}\,,\\
h^{(1)}_{qg,4} &= \frac{24409 - 69264\,x + 62052\,x^2 -
24052\,x^3 + 6855\,x^4}{1814400\,x^2}\,,\\
h^{(1)}_{qg,6} &= \frac{-181104 + 690137\,x -
  1008064\,x^2 + 658284\,x^3 - 214280\,x^4 + 55027\,x^5}{108864000\,x^3}\,.
\label{eq::hqg1}
\end{split}
\end{equation}

And finally, for the $q\bar q$ channel, it is
\begin{equation}
\begin{split}
h^{(1)}_{q\bar q,0} &= \frac{32}{27}\,(1 - x)^3\,,\qquad h^{(1)}_{q\bar
  q,2} = \frac{88}{405}\,\frac{(1 - x)^3}{x}\,,\qquad h^{(1)}_{q\bar q,4} =
\frac{(1 - x)^3\,(3487 + 842\,x)}{85050\,x^2}\,,\\
 h^{(1)}_{q\bar q,6} &=
\frac{(1 - x)^3\,(41160 + 16271\,x + 5573\,x^2)}{5103000\,x^3}\,.
\label{eq::hqqbar1}
\end{split}
\end{equation}
Note that except for the $1/\mtop^2$ terms for the $gg$ channel which
can be found in Ref.\,\cite{Dawson:1993qf}, even this \nlo{} expansion
has never been given in the literature. For the sake of brevity, we
quote only the first four terms in $1/\mtop^2$ here; higher order terms
are available from the authors upon request. The same is true for the
result for general values of $\muF$ and $\muR$.

There are several observations to be made: 
\begin{itemize}
\item Pulling out the leading order top mass dependence in terms of
$\sigma_0$, as done in \eqn{eq::sigmahat}, absorbs all the
logarithmic $x$ and $(1-x)$ dependence into the leading $1/\mtop$ terms.
\item In general, each power in $\mhiggs^2/\mtop^2$ is accompanied by a
  power in $1/x$. This is a consequence of assuming that the top quark mass
  is the heaviest scale of the process. The expansion therefore is not
  just in $\mhiggs^2/\mtop^2$ as suggested by the form of
  \eqn{eq::massexp}, but also in $\hat s/\mtop^2$ which leads to
  \begin{equation}
    \begin{split}
      \frac{\hat s}{\mtop^2} = \frac{\mhiggs^2}{\mtop^2}\frac{1}{x}\,.
    \end{split}
  \end{equation}
  These terms are a sign of the breakdown of the heavy-top limit: for
  large $\hat s$, they lead to a singular behaviour of the partonic
  cross section that becomes stronger with every order in
  $1/\mtop^2$. The consequence is that the corresponding hadronic cross
  section does not converge as successively higher orders in $1/\mtop{}$
  an included. Our solution of this problem will be described below.
\item For the $gg$ channel, the coefficient of the $1/x$ term at $i=2$
  turns out to vanish. The strategy of Ref.\,\cite{Dawson:1993qf} for
  analysing the $1/\mtop{}$ corrections at \nlo{} was therefore to
  discard the terms with $i > 2$ in the $gg$ channel, and to replace the
  $1/\mtop$ expansion for the other channels by the exact result (which
  is much easier to obtain than for the $gg$ channel). For the \nnlo{}
  case, we do not have that option, because the calculation of the full
  mass dependence is currently out of reach. As mentioned before, we
  will discuss the problem of low $x$ (large $\hat s$) in more detail in
  Section~\ref{sec::smallx}.
\end{itemize}

Let us now turn to the \nnlo{} results. For the soft+virtual terms,
we obtain:

\begin{equation}
\begin{split}
a^{(2)}_0 &=
\frac{11399}{144} + \frac{19}{8}\lht + \frac{133}{2}\zeta_2 
 - \frac{165}{4}\zeta_3 - \frac{9}{8}\zeta_4
+ n_l\left(-\frac{1189}{144} + \frac{2}{3}\lht - \frac{5}{3}\zeta_2 +
\frac{5}{6}\zeta_3\right)\,,\\
a^{(2)}_{2} &= 
-\frac{47437199}{1244160} 
+\zeta_2\left( \frac{89}{45}
+ \frac{7}{45}\ln 2\right)
+ \frac{1909181}{55296}\,\zeta_3
+ \frac{883}{1080}\,\lht 
\\&\qquad
+ n_l\,\bigg(\frac{14563}{48600} 
- \frac{281}{2880}\,\lht 
- \frac{7}{90}\,\zeta_2
\bigg) \,,\\
a^{(2)}_{4} &= 
-\frac{998645169149}{117050572800} 
+ \zeta_2\,\left(\frac{9677}{37800} 
+ \frac{857}{37800}\,\ln 2\right)
+ \frac{267179777}{35389440}\,\zeta_3
+ \frac{4039}{51840}\,\lht 
\\&\qquad
+ n_l\,\left(\frac{4565713}{285768000} 
  - \frac{857}{75600}\,\zeta_2
  - \frac{193927}{21772800}\,\lht 
\right) \,,\\
a^{(2)}_{6} &=
-\frac{1712964005499545249}{39328992460800000} 
 + \zeta_2\left(\frac{646571}{15876000} +
   \frac{17881}{4536000}\ln 2 \right)
+  \frac{5756378217151}{158544691200}\,\zeta_3
\\&\qquad
+ \frac{88077779}{7620480000}\,\lht 
+ n_l\,\left(\frac{8432587511}{4800902400000} 
-    \frac{17881}{9072000}\,\zeta_2
-\frac{111726613}{91445760000}\,\lht 
\right)
\,,\\[1em]
b^{(2)}_{0,0} &= 
 -\frac{101}{3} 
+ 33\,\zeta_2 
+ \frac{351}{2}\,\zeta_3
+ n_l\,\left(
    \frac{14}{9} 
   - 2\,\zeta_2\right)\,,\\
b^{(2)}_{1,0} &=  
  133 
- 90\,\zeta_2
- \frac{10}{3}\,n_l \,,\qquad
b^{(2)}_{1,2} = \frac{136}{45}\,,\qquad
b^{(2)}_{1,4} = \frac{3553}{9450}\,,\qquad
b^{(2)}_{1,6} =  \frac{917641}{15876000}\,,\\
b^{(2)}_{2,0} &= -33 + 2\,n_l\,,\qquad
b^{(2)}_{2,i} = 0\quad\forall\ i\geq 2\,,\\
b^{(2)}_{3,0} &= 72\,,\qquad
b^{(2)}_{3,i} = 0\quad\forall\ i\geq 2\,,
\label{eq::soft2}
\end{split}
\end{equation}
where $\zeta_n\equiv \zeta(n)$ is Riemann's zeta function with values
\begin{equation}
\begin{split}
\zeta_2 = \frac{\pi^2}{6} = 1.64493\ldots\,,\qquad 
\zeta_3 = 1.20206\ldots\,,\qquad
\zeta_4= \frac{\pi^4}{90} = 1.08232\ldots\,,
\end{split}
\end{equation}
and
\begin{equation}
\begin{split}
\lht = \ln\frac{\mhiggs^2}{\mtop^2}\,.
\end{split}
\end{equation}

The various hard contributions are evaluated as expansions around $x=1$.
We quote them through $\order{1-x}$ for the sake of brevity. Terms
through order $(1-x)^{13}$, for arbitrary values of $\muF$ and $\muR$,
are available upon request from the authors.

For the $gg$ channel, we find

\begin{equation}
\begin{split}
h^{(2)}_{gg,0} &= \frac{1453}{12} - 147\,\zeta_2 - 351\,\zeta_3 +
\lnxm{}\,\bigg(-\frac{1193}{4} + 180\,\zeta_2\bigg) +
\frac{411}{2}\,\lnxm{2} - 144\,\lnxm{3} \\&\qquad + n_l\,\bigg(
-\frac{77}{18} + 4\,\zeta_2 + \frac{101}{12}\,\lnxm{} - 4\,\lnxm{2}
\bigg) \\& +(1-x)\bigg[ -\frac{3437}{4} + \frac{1017}{2}\,\zeta_2 +
  \frac{1053}{2}\,\zeta_3 + \lnxm{}\,\bigg(\frac{2379}{2} -
  270\,\zeta_2\bigg) \\&\qquad - \frac{2385}{4}\,\lnxm{2} +
  216\,\lnxm{3} + n_l\,\bigg( \frac{395}{24} - \frac{22}{3}\,\zeta_2 -
  \frac{45}{2}\,\lnxm{} + \frac{22}{3}\,\lnxm{2} \bigg) \bigg] + \cdots\,,
\\ h^{(2)}_{gg,2} &= \frac{68}{45} - \frac{272}{45}\,\lnxm{} +
(1-x)\bigg[ -\frac{6661}{1200} - \frac{9}{80}\,\zeta_2 +
  \frac{172}{15}\,\lnxm{} - \frac{81}{80}\,\lnxm{2} -
  \frac{27}{40}\,\lht \\&\qquad + n_l\,\bigg( \frac{11}{80}
  -\frac{1}{20} \lnxm{} \bigg)\bigg] + \cdots\,,\\ h^{(2)}_{gg,4} &=
\frac{3553}{18900} - \frac{3553}{4725}\,\lnxm{} + (1-x)\bigg[
  -\frac{1837333}{10584000} + \frac{21}{3200}\,\zeta_2 +
  \frac{49927}{50400}\,\lnxm{} \\&\qquad + \frac{189}{3200}\,\lnxm{2} +
  \frac{471}{11200}\,\lht + n_l\,\bigg( -\frac{659}{67200} +
  \frac{7}{2400}\,\lnxm{} \bigg) \bigg]+\cdots\,,\\ h^{(2)}_{gg,6} &=
\frac{917641}{31752000} - \frac{917641}{7938000}\,\lnxm{} + (1-x)\bigg[
  \frac{13461173}{635040000} + \frac{1697}{896000}\,\zeta_2 +
  \frac{5644979}{42336000}\,\lnxm{} \\&\qquad +
  \frac{15273}{896000}\,\lnxm{2} + \frac{15731}{1344000}\,\lht +
  n_l\,\bigg( - \frac{23347}{8064000} + \frac{1697}{2016000}\,\lnxm{}
  \bigg) \bigg] + \cdots\,.
\label{eq::hgg2}
\end{split}
\end{equation}

Here and in the following equations, the ellipse indicate higher orders
in $(1-x)$.  The $qg$ (and $\bar qg$) channel reads

\begin{equation}
\begin{split}
h^{(2)}_{qg,0} &=
\frac{11}{27} 
+ \frac{29}{6}\,\zeta_2 
+ \frac{311}{18}\,\zeta_3
+ \frac{85}{36}\,\lnxm{2} 
+ \frac{367}{54}\,\lnxm{3} 
\\&\qquad
+ n_l\,\bigg(
  \frac{13}{81} 
  - \frac{2}{3}\,\lnxm{} 
  +\frac{1}{18}\,\lnxm{2}
\bigg) 
+ \lnxm{}\,\bigg(
  \frac{341}{18} 
  - \frac{50}{9}\,\zeta_2
\bigg)
\\&\qquad 
+ (1-x)\bigg[
-\frac{959}{18} 
+ 8\,\zeta_2
+ \frac{433}{9}\,\lnxm{} 
- \frac{33}{2}\,\lnxm{2} 
+ \frac{4}{9}\,n_l\,\lnxm{}
\bigg]+\cdots\,,\\
h^{(2)}_{qg,2} &=
\frac{68}{405} 
+ \frac{136}{405}\,\lnxm{}
+ (1-x)\bigg[
-\frac{62737}{24300} 
- \frac{539}{1080}\,\zeta_2
+ \frac{4367}{3240}\,\lnxm{} 
- \frac{187}{360}\,\lnxm{2} 
\\&\qquad
- \frac{1441}{3240}\,\lht 
+ n_l\,\bigg(
  \frac{44}{405} 
  - \frac{11}{270}\,\lnxm{}\bigg)
\bigg] + \cdots\,,\\
h^{(2)}_{qg,4} &=
\frac{3553}{170100} 
+ \frac{3553}{85050}\,\lnxm{}
+ (1-x)\bigg[
-\frac{66227323}{285768000} 
- \frac{2947}{129600}\,\zeta_2
+ \frac{26401}{388800}\,\lnxm{} 
\\&\qquad
- \frac{7157}{302400}\,\lnxm{2} 
- \frac{37481}{2721600}\,\lht 
+ n_l\,\bigg(
  \frac{421}{85050} 
  - \frac{421}{226800}\,\lnxm{}\bigg) 
\bigg]+\cdots\,,\\
h^{(2)}_{qg,6} &=
\frac{917641}{285768000} 
+ \frac{917641}{142884000}\,\lnxm{}
+ (1-x)\bigg[
-\frac{1018432391}{34292160000} 
\\&\qquad
- \frac{39011}{15552000}\,\zeta_2
+ \frac{2460599}{326592000}\,\lnxm{} 
- \frac{94741}{36288000}\,\lnxm{2} 
- \frac{96389}{65318400}\,\lht 
\\&\qquad
+ n_l\,\bigg(
  \frac{5573}{10206000} 
  - \frac{5573}{27216000}\,\lnxm{}
\bigg) 
\bigg] + \cdots\,.\\
\end{split}
\end{equation}

The remaining channels only start at higher orders in $(1-x)$:

\begin{equation}
\begin{split}
h^{(2)}_{q\bar q,0} =h^{(2)}_{qq,0} = h^{(2)}_{qq',0} &= (1-x)\bigg[
\frac{20}{9} 
- \frac{16}{9}\,\zeta_2
- \frac{16}{9}\,\lnxm{} 
+ \frac{16}{9}\,\lnxm{2} 
\bigg] + \cdots\,,\\
h^{(2)}_{q\bar q,i} =h^{(2)}_{qq,i} = h^{(2)}_{qq',i} &=
\order{(1-x)^2}\quad
\forall\ i\geq 2\,.
\label{eq::hqq2}
\end{split}
\end{equation}

Again, we have included the known leading terms in $1/\mtop$ for the
sake of completeness.

\section{Small-$x$ behavior}\label{sec::smallx}

As pointed out above, the $1/\mtop$ expansion cannot give the proper
result in the low-$x$ (large-$\hat s$) region, which is why the
$(1-x)$-expansion in this approach is no worse than the full $x$
dependence.  Fortunately, in Ref.\,\cite{Marzani:2008az}, it was derived
that the leading small-$x$ behavior of the partonic cross section is
given by ($\muF=\muR=\mhiggs$)
\begin{equation}
\begin{split}
\hat\sigma^{(1)}_{gg}(x) &= 3\,\sigma_0\,{\cal C}^{(1)}+ \order{x}\,,\qquad
\hat\sigma^{(2)}_{gg}(x) = -9\,\sigma_0\,{\cal C}^{(2)}\ln x + c + \order{x}\,,
\label{eq::xto0}
\end{split}\end{equation}
where the coefficients ${\cal C}^{(1)}$ and ${\cal C}^{(2)}$ are
available in Ref.\,\cite{Marzani:2008az} in the form of a numerical
table for various values of $\mtop/\mhiggs$ out of which we constructed
simple interpolating functions.  The constant $c$ was undetermined.
Note that \eqn{eq::xto0} is the actual limit of $\hat s\to \infty$,
i.e., it is {\it not} derived in the heavy-top limit.

\begin{figure}
  \begin{center}
    \begin{tabular}{cc}
      \includegraphics[bb=110 265 465
        560,width=.45\textwidth]{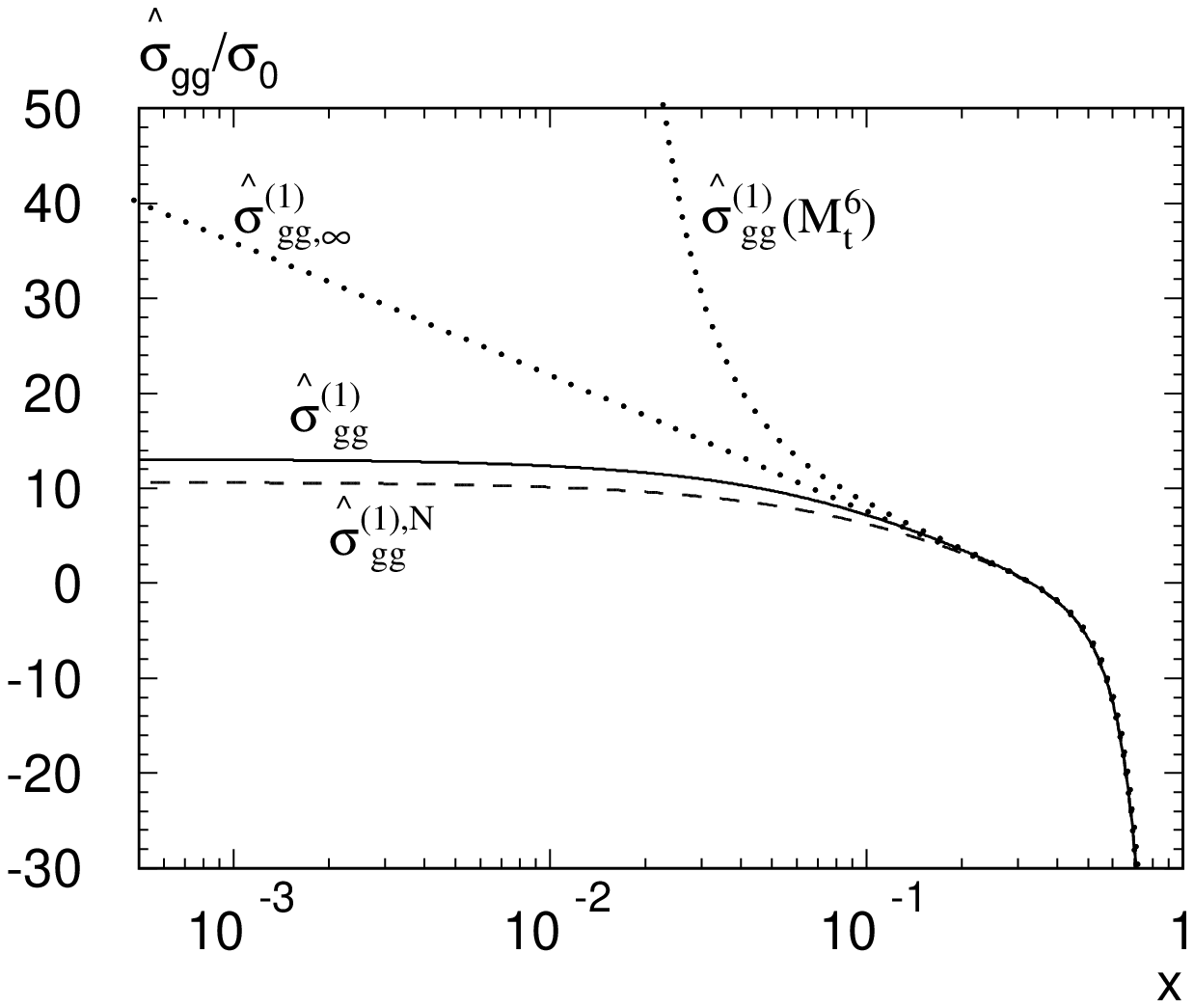}
      &
      \includegraphics[bb=110 265 465
        560,width=.45\textwidth]{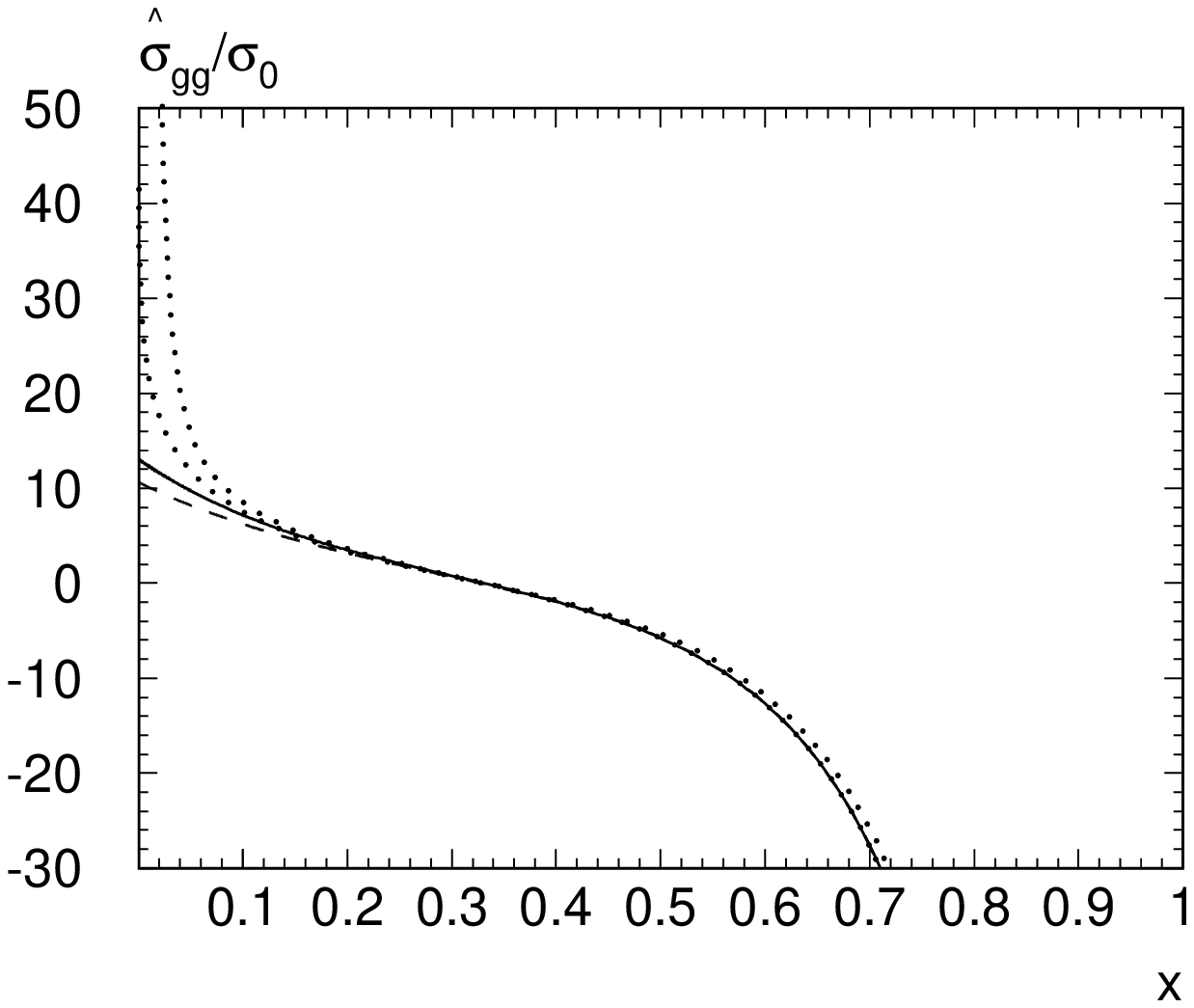}\\
      (a) & (b)
    \end{tabular}
    \parbox{.9\textwidth}{
      \caption[]{\label{fig::marzani1}\sloppy The \nlo{} contribution to
        the partonic cross section $\hat\sigma^{(1)}_{gg}$ as a function
        of $x=\mhiggs^2/\hat s$ on a logarithmic (a) and a linear scale
        (b). The expansion in $(1-x)$ converges up to the threshold
        $x=\mhiggs^2/(4\mtop^2)\approx 0.14$. This expansion, through
        $(1-x)^{8}$, is displayed as the dashed curve,
        $\hat\sigma_{gg}^{(1),N}$, $N=8$. The full $x$-dependence of the
        $1/\mtop{}$ expansion is shown as the dotted curves (lower:
        leading term in $\mhiggs/\mtop$, upper: including terms of order
        $1/\mtop^6$). The solid line is the combination of the
        $(1-x)$-expansion with the leading behavior at $x\to 0$,
        cf.~\eqn{eq::xto0}. }}
  \end{center}
\end{figure}

We may use this additional information to improve our result in
the following way:\footnote{ The upper equation essentially corresponds to the
matching procedure suggested in Ref.\,\cite{Marzani:2008az}.}
\begin{equation}
\begin{split}
\hat\sigma^{(1)}_{gg}(x) &= \hat\sigma^{(1),N}_{gg}(x) +
(1-x)^{N+1}\,\left[3\,\sigma_0 {\cal C}^{(1)} - \hat\sigma^{(1),N}_{gg}(0)
  \right]\,,\\ \hat\sigma^{(2)}_{gg}(x) &= \hat\sigma^{(2),N}_{gg}(x) -
9\,\sigma_0{\cal C}^{(2)}\left[ \ln x + \sum_{k=1}^N\frac{1}{k}(1-x)^k
  \right]\,,
\label{eq::match}
\end{split}
\end{equation}
where $\hat\sigma_{gg}^{(n),N}$ denotes the expansion of the partonic
cross section around $x=1$ through $\order{(1-x)^N}$. Note that
$\hat\sigma^{(1)}_{gg}(x)$ and $\hat\sigma^{(2)}_{gg}(x)$ have the
correct behavior for $x\to 0$ and $x\to 1$ up to the orders
considered. In this way, we arrive at smooth functions that approximate
the full partonic cross section over the full $x$-range. In order to
illustrate the quality of this method, \fig{fig::marzani1} shows
$\hat\sigma^{(1)}_{gg}(x)$ together with the soft expansion
$\hat\sigma^{(1),N}_{gg}(x)$ for $N=8$, as well as
$\hat\sigma^{(1)}_{gg,\infty}$, i.e., the full $x$-dependence of the
heavy-top result.  All expressions include top mass corrections through
$\order{1/\mtop^6}$; the curves are normalized by $\sigma_0$.

In order to be able to directly compare \fig{fig::marzani1} with
Ref.\,\cite{Marzani:2008ih} (which updates the numerics of
Ref.\,\cite{Marzani:2008az}), the scales in \fig{fig::marzani1}~(a) are
chosen identical to those of Figs.\,1 and 2 in
Ref.\,\cite{Marzani:2008ih}.  For the same reason, we set
$\mtop=170.9$\,GeV at this point, in contrast to the actual numerical
section of this paper where the current world average for $\mtop$ is
used. And finally, the curve for $\hat\sigma^{(1)}_{gg,\infty}$ in the
heavy-top limit is included (i.e., only the $1/\mtop^0$ terms), as in
Ref.\,\cite{Marzani:2008ih}. In case one is misled by the apparently
large effect, \fig{fig::marzani1}~(b) shows the same curves on a linear
scale. In fact, as will become obvious shortly, the effect of this
matching on the {\it hadronic} cross section is rather small compared to
using just the pure soft expansion $\hat\sigma^{(1),N}_{gg}(x)$
(cf.~\fig{fig::ratnloggmt} below).

Similarly, \fig{fig::marzani2} shows the \nnlo{} partonic cross section
(gluon-gluon channel) as constructed from \eqn{eq::match}, again with
choice of scales and set of parameters as in
Ref.\,\cite{Marzani:2008ih}.  The agreement to the result obtained in
Ref.\,\cite{Marzani:2008ih} is good, with a small difference on the
unknown constant $c$ in \eqn{eq::match}. The effect of the matching on
the {\it hadronic} cross section is completely negligible when compared
to the pure soft expansion $\hat\sigma^{(2),N}_{gg}(x)$
(cf.~\fig{fig::ratnnloggmt} below). This is of course due to a
suppression of the large-$x$ region by the parton densities.

\begin{figure}
  \begin{center}
    \begin{tabular}{cc}
      \includegraphics[bb=110 265 465
        560,width=.45\textwidth]{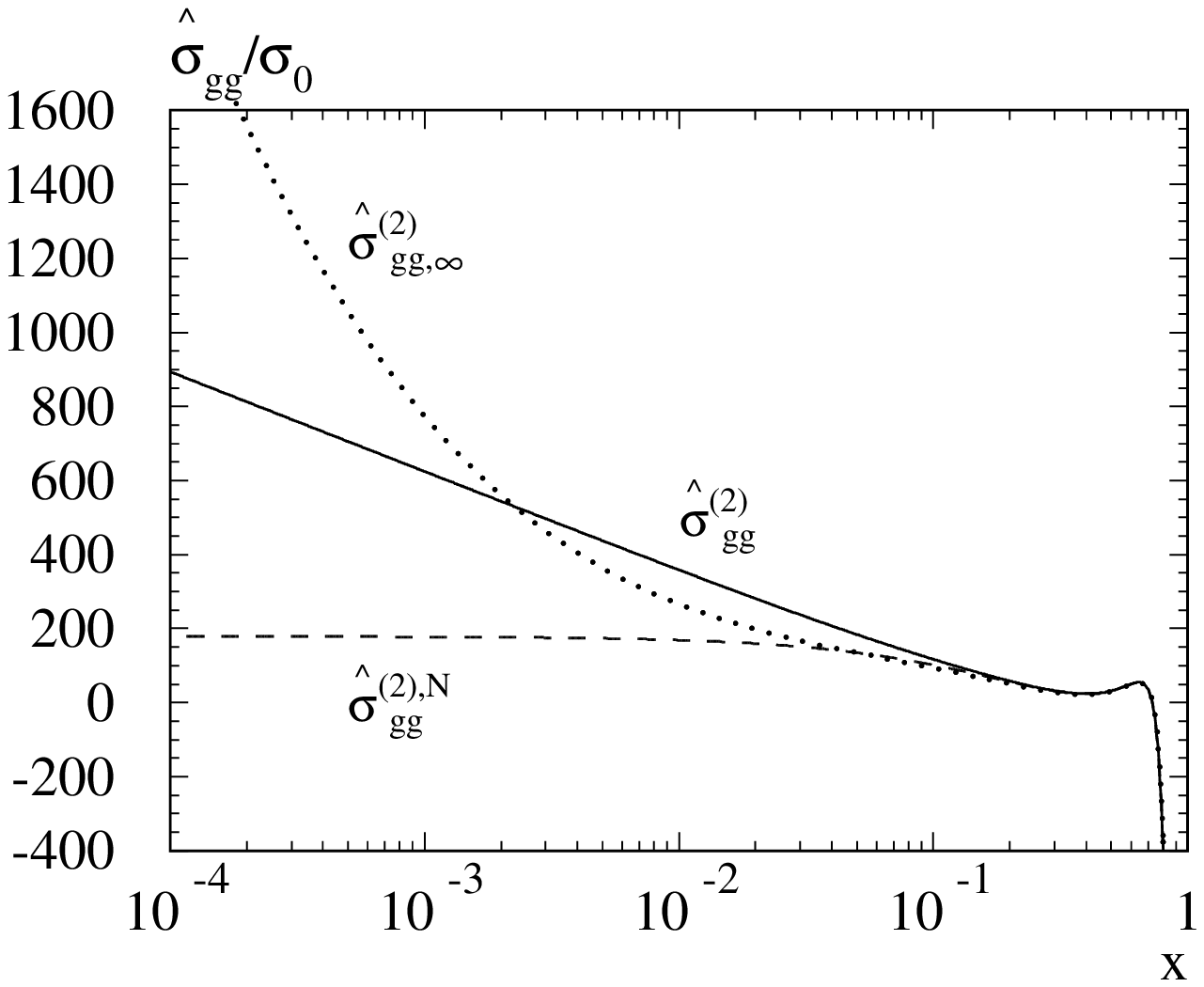} &
      \includegraphics[bb=110 265 465
        560,width=.45\textwidth]{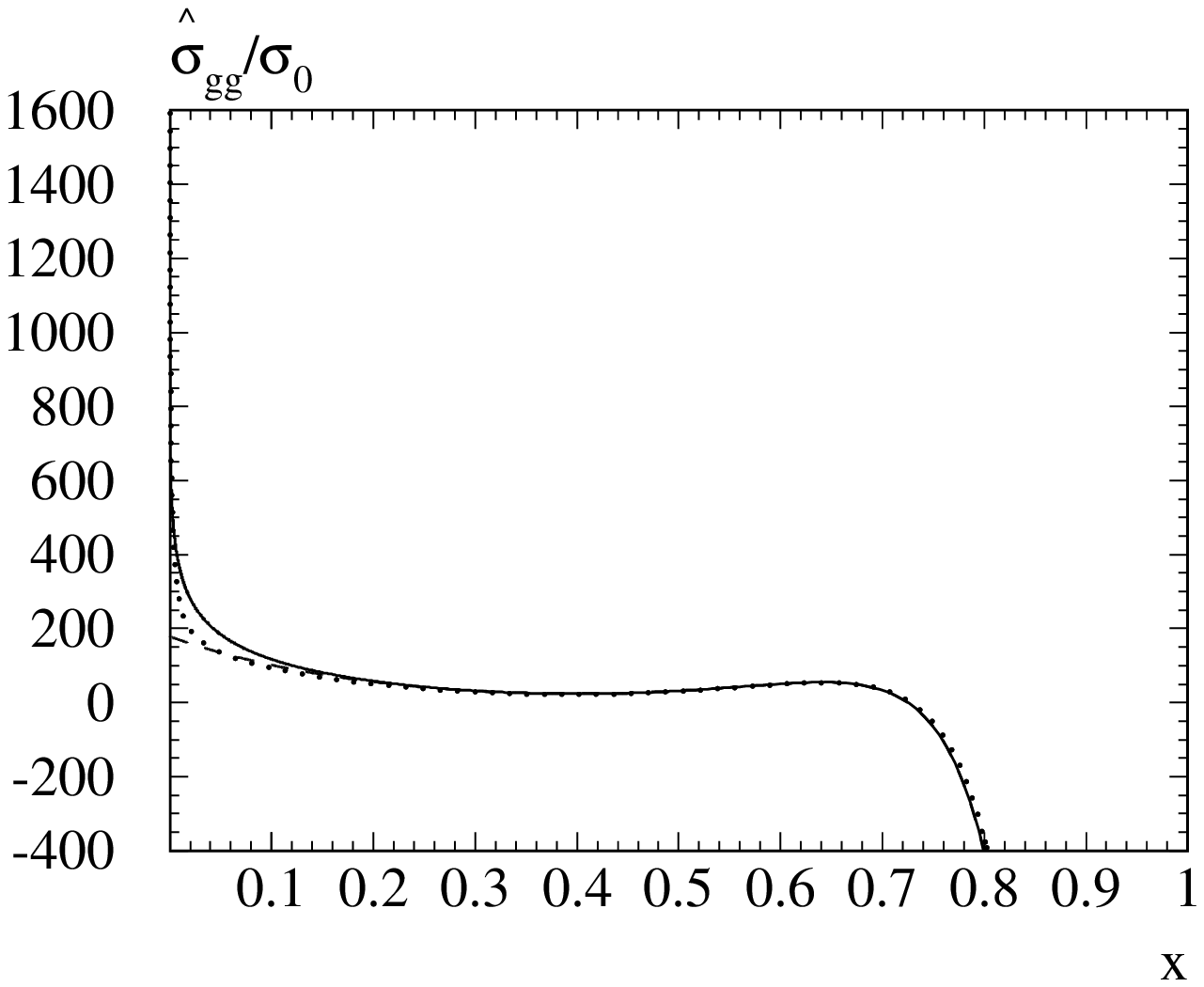}\\
      (a) & (b)
    \end{tabular}
    \parbox{.9\textwidth}{
      \caption[]{\label{fig::marzani2}\sloppy
        Same as \fig{fig::marzani1}, but at \nnlo{}. The dotted line is
        the full $x$-dependence of the $1/\mtop^0$ result.
        }}
  \end{center}
\end{figure}

\section{Numerical results}\label{sec::numerics}

Unless stated otherwise, in all our numerical analyses in this paper, we
use $\mtop = 173.1$\,GeV and set $\muF=\muR=\mhiggs$. The latter
restriction shall be sufficient for this first study of the top mass
effects; more detailed phenomenological studies, including scale
variations, are left for a future publication.

\subsection{Next-to-leading order}

The natural extension of the heavy top limit of \eqn{eq::heavytop} would
be to use \eqn{eq::sigmahat} with the full top mass dependence in
$\sigma_0$ and the $1/\mtop{}$ expansion of $\Delta_{\alpha\beta}$ as
given in Section~\ref{sec::anares}, and to match the result to the
large-$\hat s$ region as defined in Section~\ref{sec::smallx}. However,
in order to strictly test the heavy-top limit, we prefer to apply a
consistent $1/\mtop{}$ expansion to the partonic cross section, without
factoring out the \lo{} mass dependence. Once the convergence of this
$1/\mtop{}$ expansion and its consistency with the heavy-top limit of
\eqn{eq::heavytop} is shown, one could try to derive an ``improved
heavy-top limit'' by keeping the full mass dependence in $\sigma_0$.
However, as we will see, the improvement achieved in this way is well
below any expected experimental accuracy.

At \nlo{}, we therefore define
\begin{equation}
\begin{split}
\hat\sigma^\nlo_{\alpha\beta}(\mtop^n) &=
\sigma_0\,\delta_{\alpha g}\delta_{\beta g}\delta(1-x)
+ \api\, \hat\sigma_{\alpha\beta}^{(1)}(\mtop^n)\,,
\label{eq::signlomtexp}
\end{split}
\end{equation}
where $\hat\sigma_{\alpha\beta}^{(1)}(\mtop^n)$ is the \nlo{}
contribution to the partonic cross section evaluated as an expansion
through $\order{(\mhiggs/\mtop)^n}$. It is obtained by expanding
$\sigma_0\Delta^{(1)}_{\alpha\beta}$ with $\sigma_0$ and
$\Delta^{(1)}_{\alpha\beta}$ from Eqs.\,(\ref{eq::sigma0}),
(\ref{eq::softhard}), (\ref{eq::soft1})--(\ref{eq::hqqbar1}) in terms of
$1/\mtop$, and applying the matching procedure of \eqn{eq::match}. The
corresponding hadronic quantity derived from \eqn{eq::signlomtexp} is
denoted by $\sigma^\nlo_{\alpha\beta}(\mtop^n)$, as usual. Note that it
also depends on the depth of the expansion in $(1-x)$.

\begin{figure}
  \begin{center}
    \begin{tabular}{cc}
      \includegraphics[bb=110 265 465
        560,width=.45\textwidth]{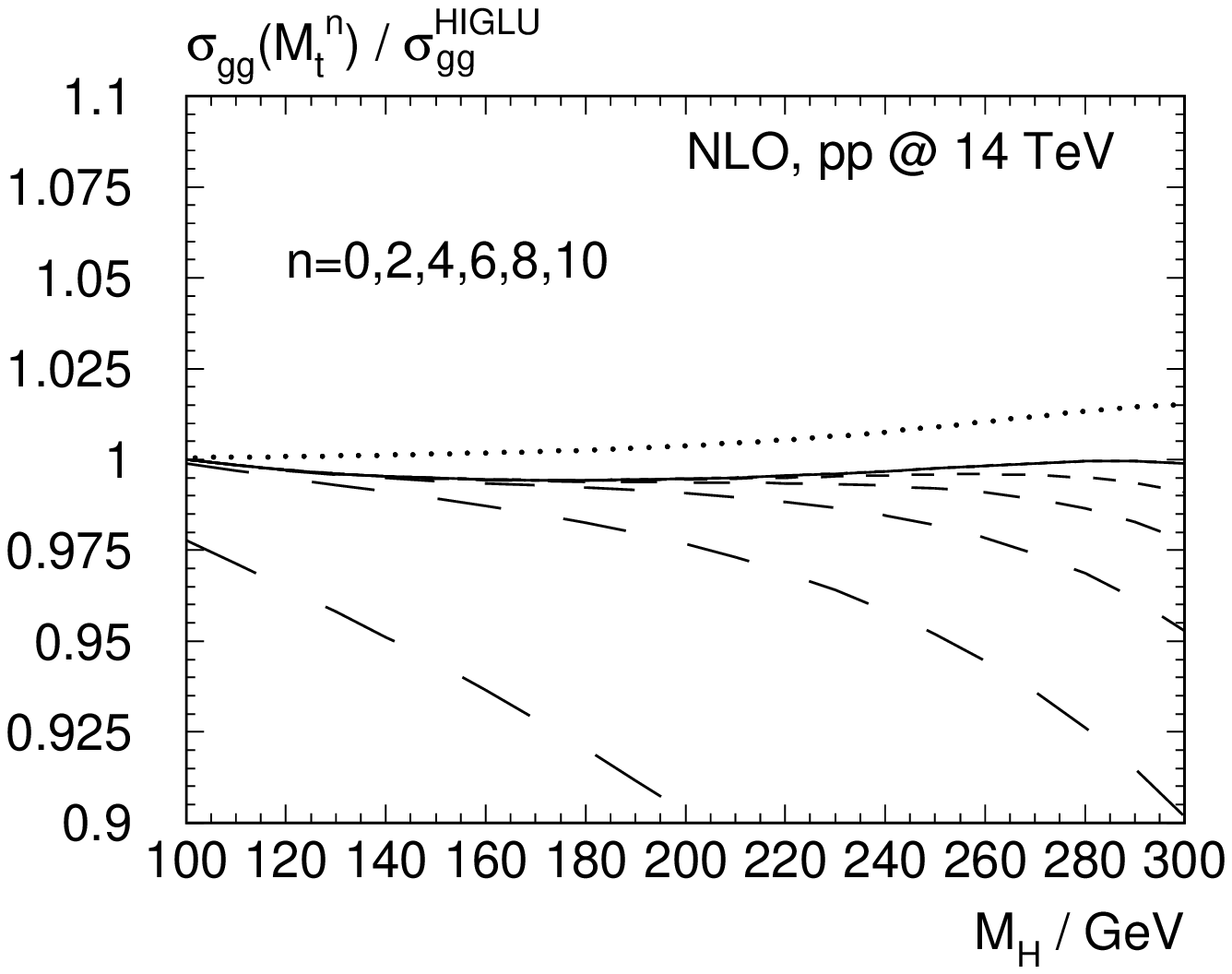} &
      \includegraphics[bb=110 265 465
        560,width=.45\textwidth]{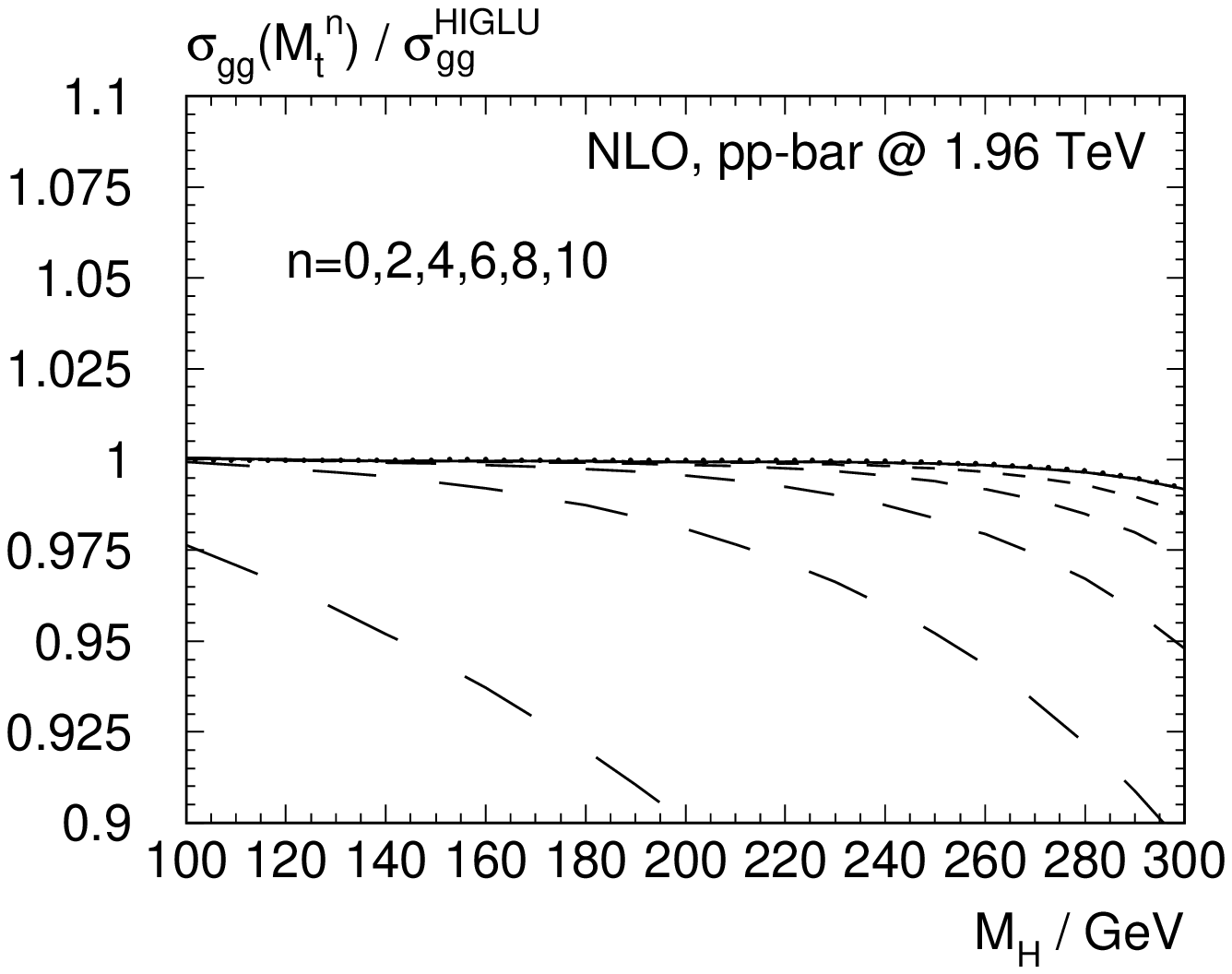}\\
      (a) & (b)
    \end{tabular}
    \parbox{.9\textwidth}{
      \caption[]{\label{fig::ratnloggmt}\sloppy Ratio of the $gg$
        induced component of the \nlo{} hadronic cross section as
        obtained from \eqn{eq::match} to the value obtained from {\tt
          HIGLU}~\cite{Spira:1995mt}, when keeping successively higher
        orders in $1/\mtop$ (decreasing dash-length corresponds to
        increasing order); the dotted line is the result obtained from
        the pure soft expansion $\hat\sigma_{gg}^{(1),N}$ through order
        $1/\mtop{}^{10}$ without the matching of \eqn{eq::match}.}}
  \end{center}
\end{figure}

First, we look at the convergence of the $1/\mtop$ expansion of the $gg$
channel alone, whose low-$x$ behaviour is implemented as described in
Section~\ref{sec::smallx}. \fig{fig::ratnloggmt} shows the ratio of
$\sigma^\nlo_{gg}(\mtop^n)$, keeping various orders in $1/\mtop$, to the
fully mass dependent result $\sigma^{\rm HIGLU}_{gg}$ (dashed:
$1/\mtop^n$, $n=0,\ldots,8$; solid: $1/\mtop^{10}$). The dotted line
corresponds to the pure soft expansion result $\sigma^N_{gg}$, without
matching to the low-$x$ behaviour. The convergence towards the exact
result is excellent, both for the \lhc{} (a) and the Tevatron (b). The
slightly better behaviour for the Tevatron is due to the smaller
high-energy region. The effect of the matching from
Section~\ref{sec::smallx} is rather small.

Unfortunately, the analytic low-$x$ behaviour is currently not known for
the subleading channels ($qg$, $q\bar q$, and also the \nnlo{} channels
$qq$, $qq'$). However, their numerical contribution at \nlo{} is very
small (for ``reasonable'' values of $\muF\sim \mhiggs$): for Higgs
masses between 100 and 300\,GeV, the $qg$ contribution is at the few
percent level ($< 4$\%) and the $q\bar q$ at the permille level.

One could therefore just proceed by ignoring the $1/\mtop{}$-corrections
at \nnlo{} for all but the $gg$ channel.  At \nlo{}, however, we observe
that we can nevertheless improve the prediction by including the
$1/\mtop$ and $(1-x)$ expansion of the subleading terms up to a certain
depth beyond which the convergence properties of the series for the
individual channels deteriorate, as is typical for an asymptotic
series. Applying this criterion at \nlo{} allows us to include the terms
through order $1/\mtop^{10}$ for the $qg$ channel, but only the first
two terms for the $q\bar q$ channel. The convergence of the expansion in
$(1-x)$ is excellent (see below).

\begin{figure}
  \begin{center}
    \begin{tabular}{cc}
      \includegraphics[bb=110 265 465
        560,width=.45\textwidth]{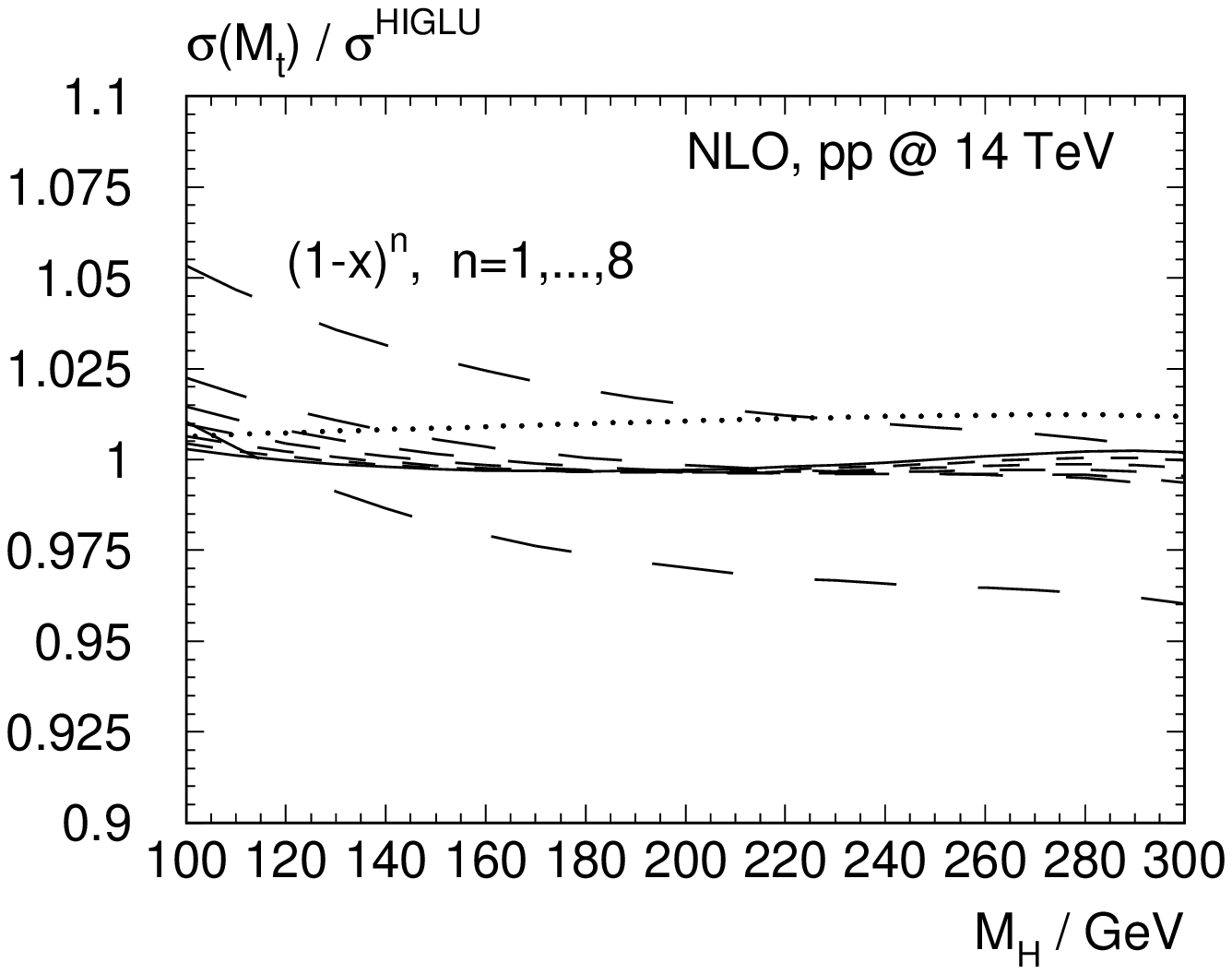} &
      \includegraphics[bb=110 265 465
        560,width=.45\textwidth]{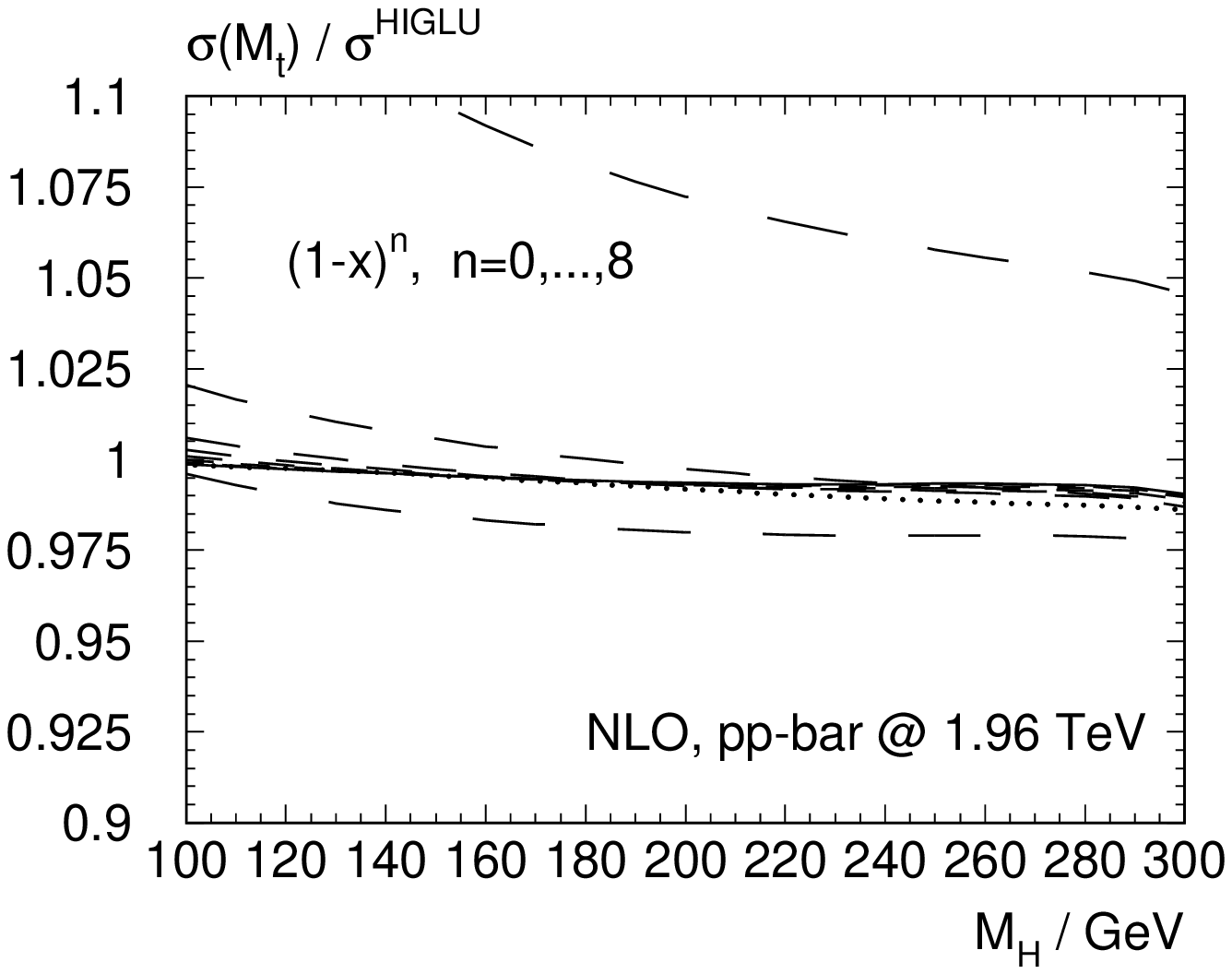}\\
      (a) & (b)
    \end{tabular}
    \parbox{.9\textwidth}{
      \caption[]{\label{fig::ratnloopt}\sloppy Solid: the final result
        for the \nlo{} cross section, divided by the full top mass
        dependent result as obtained from {\tt
          HIGLU}~\cite{Spira:1995mt}. The dashed lines correspond to
        various orders in the $(1-x)$ expansion and nicely demonstrate
        the quality of the convergence. The dotted lines represent the
        result of the heavy-top limit at \nlo{} (\eqn{eq::heavytop}). }}
  \end{center}
\end{figure}

This ``optimal'' result will be denoted by $\sigma^{\nlo}(\mtop)$. It is
shown in \fig{fig::ratnloopt}, divided by the exact mass dependence,
both for a 14\,TeV proton--proton collider (a), and for a $1.96$\,TeV
proton--anti-proton collider (b). One observes a nice convergence for
the soft expansion towards a result that reproduces the full mass
dependence at the sub-percent level.  For comparison, the figure also
contains the ratio of the heavy-top limit to the exact mass dependence
(dotted line) which deviates from one at the percent level, as pointed
out before.

\subsection{Next-to-next-to-leading order}

In analogy to \eqn{eq::signlomtexp}, we define
\begin{equation}
\begin{split}
\hat\sigma^\nnlo_{\alpha\beta}(\mtop^n) &=
\sigma_0\,\left[\delta_{\alpha g}\delta_{\beta g}\delta(1-x)
+ \api\,\Delta_{\alpha\beta,\infty}^{(1)}\right] 
+ \left(\api\right)^2\,\hat\sigma_{\alpha\beta}^{(2)}(\mtop^n)\,,
\label{eq::signnlomtexp}
\end{split}
\end{equation}
where $\hat\sigma_{\alpha\beta}^{(2)}(\mtop^n)$ is obtained by expanding
$\sigma_0\Delta^{(2)}_{\alpha\beta}$, with $\sigma_0$ and
$\Delta^{(2)}_{\alpha\beta}$ from Eqs.\,(\ref{eq::sigma0}),
(\ref{eq::softhard}), (\ref{eq::soft2})--(\ref{eq::hqq2}), in terms of
$1/\mtop$, up to power $n$, and applying the matching procedure of
\eqn{eq::match}. We already know that the first term in
\eqn{eq::signnlomtexp} is an excellent approximation to the full top
mass dependent result, and therefore \eqn{eq::signnlomtexp} provides a
suitable quantity to compare the heavy-top result of \eqn{eq::heavytop}
to the $1/\mtop$ expansion.

\begin{figure}
  \begin{center}
    \begin{tabular}{cc}
      \includegraphics[bb=110 265 465
        560,width=.45\textwidth]{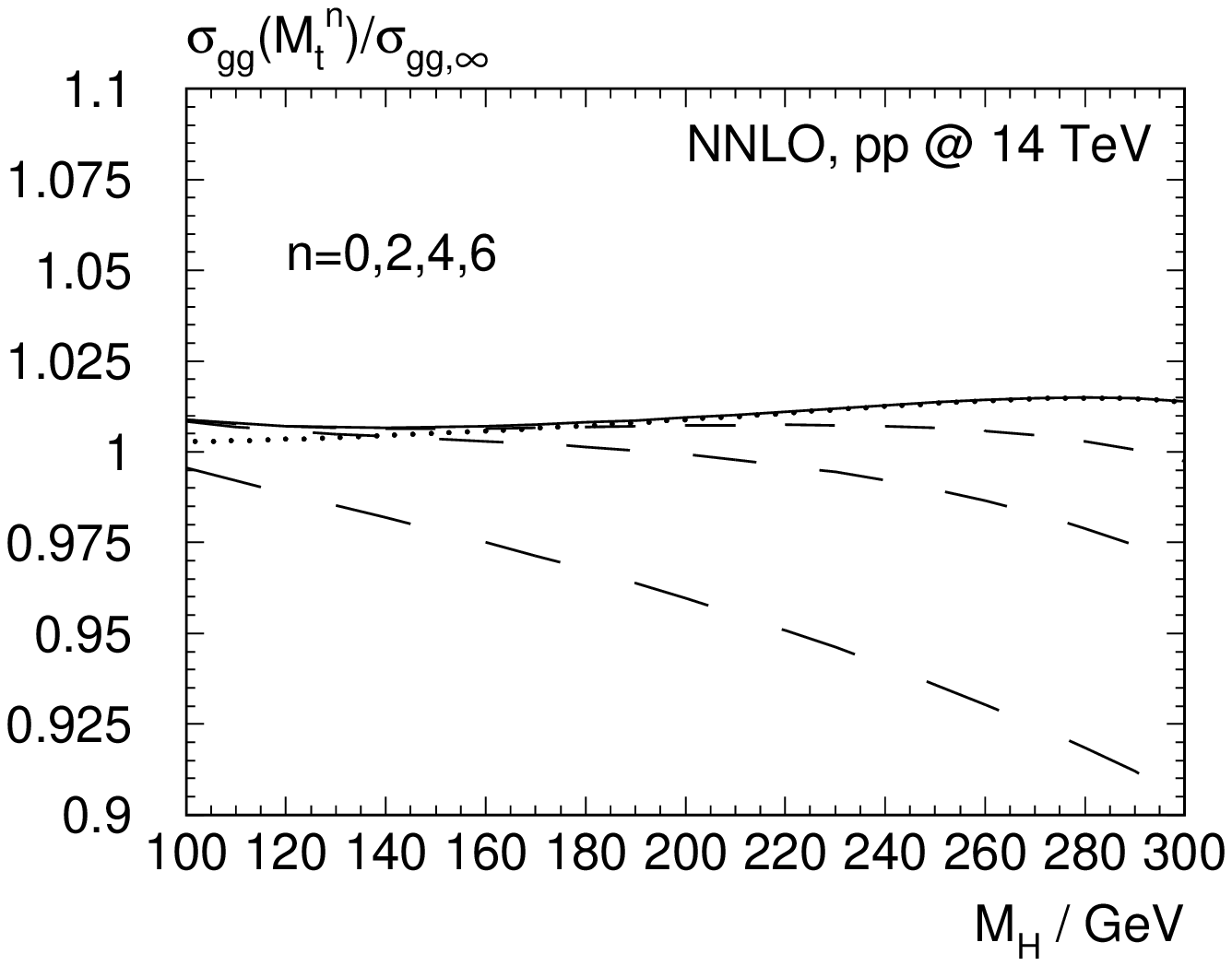} &
      \includegraphics[bb=110 265 465
        560,width=.45\textwidth]{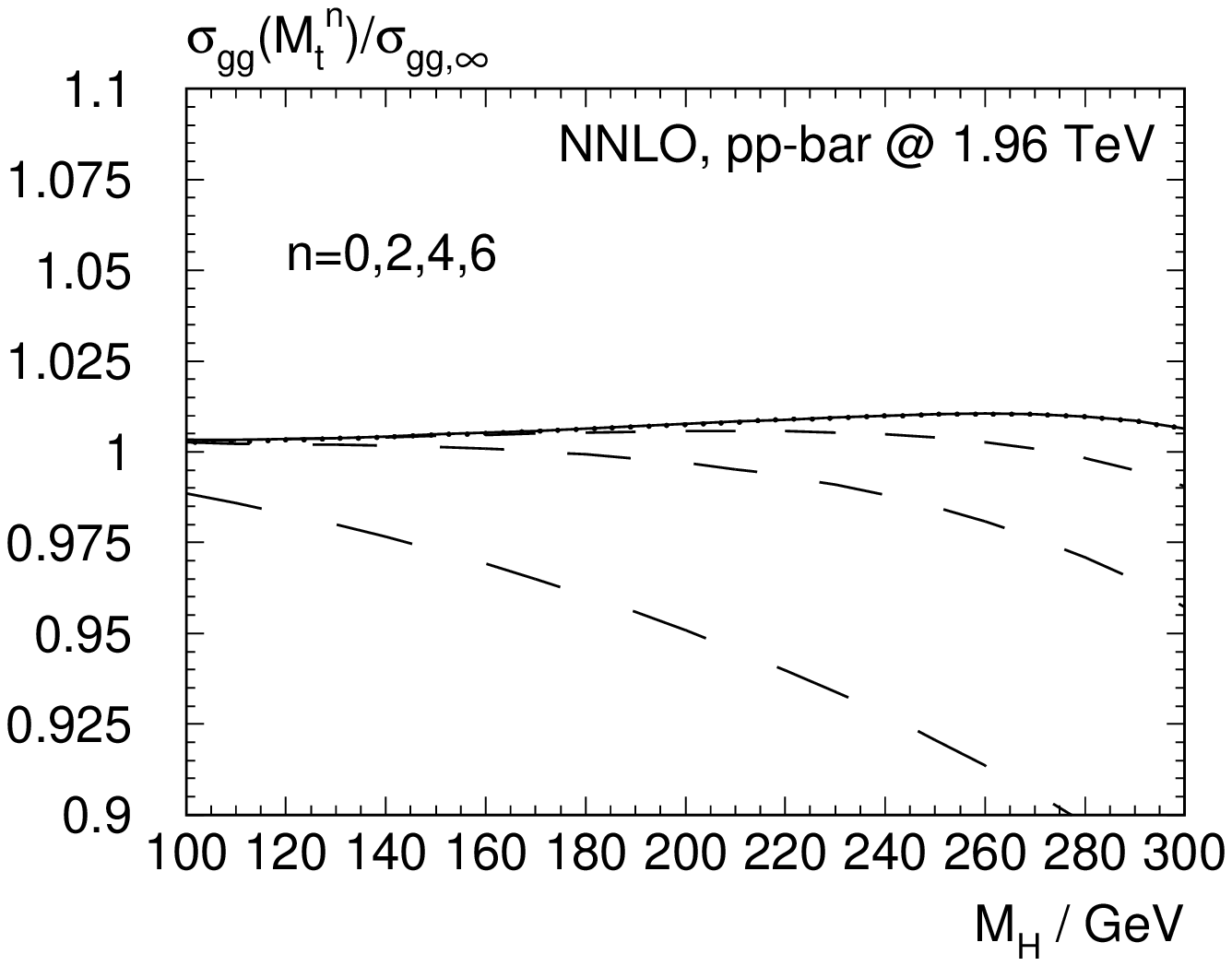}\\
      (a) & (b)
    \end{tabular}
    \parbox{.9\textwidth}{
      \caption[]{\label{fig::ratnnloggmt}\sloppy Ratio of the $gg$
        induced component of the \nnlo{} hadronic cross section as
        obtained from \eqn{eq::match} to the heavy-top result of
        \eqn{eq::heavytop}, (decreasing dash-length corresponds to
        increasing order in $1/\mtop{}$); the dotted line is the result
        obtained from the pure soft expansion $\hat\sigma_{gg}^{(2),N}$
        through order $1/\mtop{}^6$ without the matching of
        \eqn{eq::match}.}}
  \end{center}
\end{figure}

Again, we first look at the convergence of the $1/\mtop$ expansion of
the $gg$ channel alone, whose low-$x$ behaviour is implemented as
described in Section~\ref{sec::smallx}. \fig{fig::ratnnloggmt} shows the
ratio of $\sigma^\nnlo_{gg}(\mtop^n)$, keeping various orders in
$1/\mtop$, to the heavy-top result $\sigma^\nnlo_{gg,\infty}$ of
\eqn{eq::heavytop}, which we recall includes the exact LO mass
dependence (dashed: $1/\mtop^n$, $n=0,2,4$; solid: $1/\mtop^6$). The
dotted line corresponds to the pure soft expansion result
$\sigma^N_{gg}$, without matching to the low-$x$ behaviour. We observe
very good convergence towards the heavy-top result, assuring us of the
high quality of the latter.

\begin{figure}
  \begin{center}
    \begin{tabular}{cc}
      \includegraphics[bb=110 265 465
        560,width=.45\textwidth]{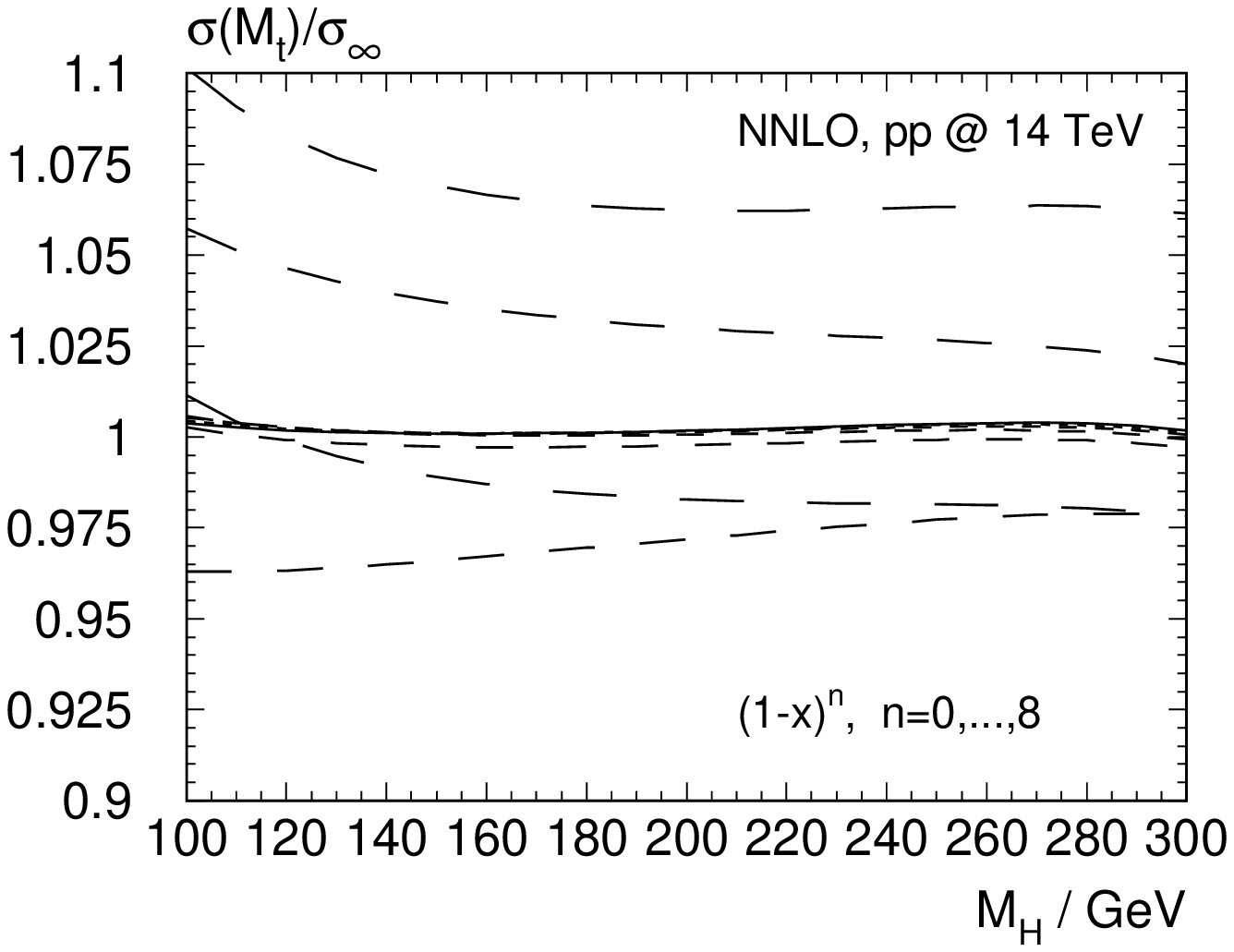} &
      \includegraphics[bb=110 265 465
        560,width=.45\textwidth]{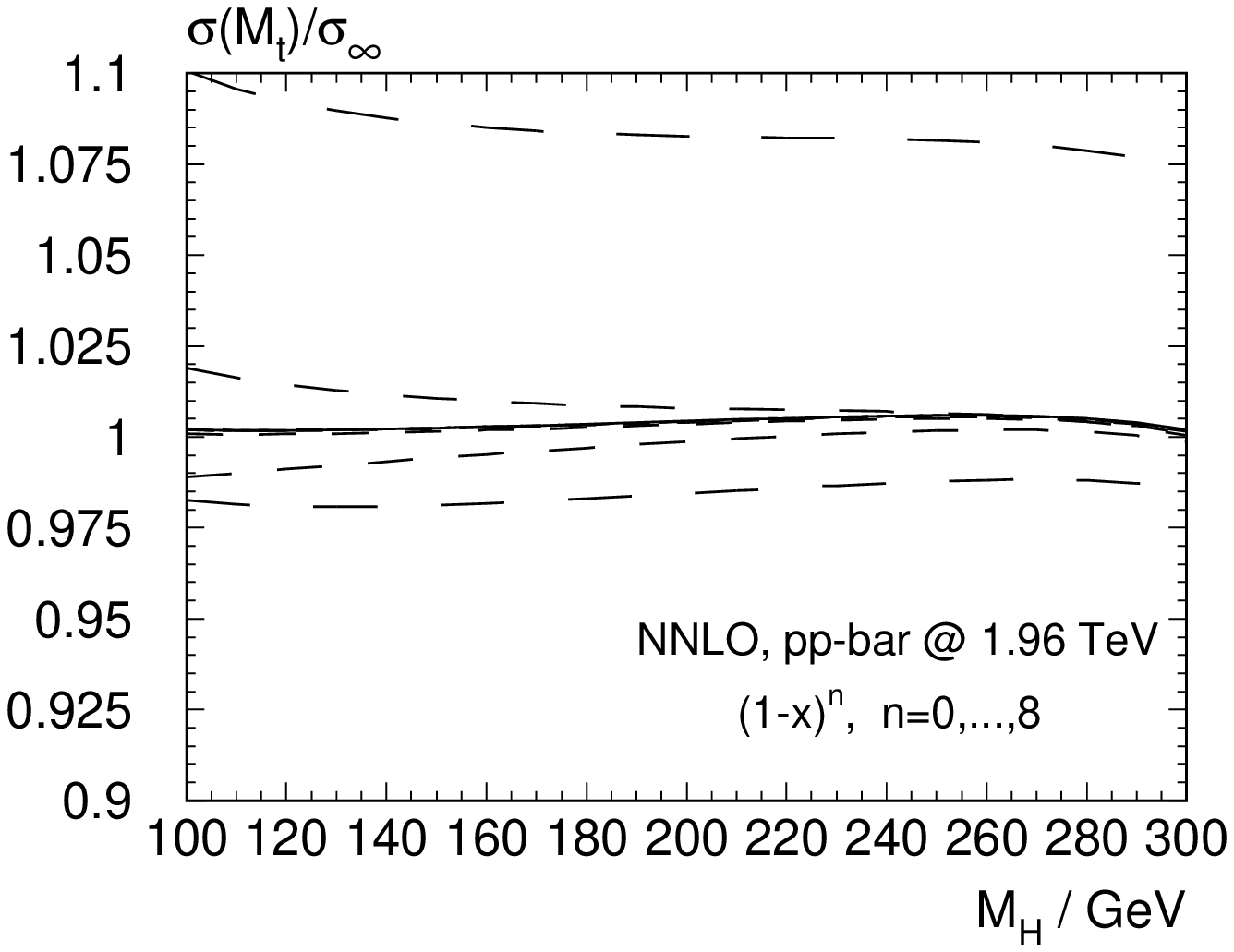}\\
      (a) & (b)
    \end{tabular}
    \parbox{.9\textwidth}{
      \caption[]{\label{fig::ratnnloopt}\sloppy \nnlo{} version of
        \fig{fig::ratnloopt}, but now the curves are normalized to the
        heavy-top limit of \eqn{eq::heavytop}.
    }}
  \end{center}
\end{figure}

We then apply the same criteria as at \nlo{} in order to obtain the
``optimal'' result $\sigma^{\nnlo}(\mtop)$ for the $\mtop$ dependent
\nnlo{} terms. They allow us to keep all four available terms for the
$gg$ and the $qg$ curve, i.e., through order $1/\mtop^6$, and again only
the first two terms for the $qq$, $q\bar q$, and $qq'$ initiated
sub-processes. The result, divided by the one obtained from
\eqn{eq::heavytop}, is shown in \fig{fig::ratnnloopt}, for various
orders in the soft expansion which again converges excellently.

The final result is within 0.5\% of the well-known heavy-top limit,
clearly justifying its successful and extensive application in the
literature.  As outlined above, one can now try to improve the \nnlo{}
prediction by factoring out the \lo{} top mass dependence. However, it
is obvious that this will not alter the result by more than 0.5\% which
is way below any expected experimental accuracy.

\section{Conclusions and Outlook}\label{sec::conclusions}

Top mass effects on the \nnlo{} Higgs cross section in gluon fusion have
been calculated. For the dominant $gg$ channel, the result for $\hat s <
4\mtop^2$ has been matched to the limiting behaviour at $x\to 0$ as
obtained in Ref.\,\cite{Marzani:2008az}. We have demonstrated the
reliability of our result through the excellent convergence of the soft
expansion. The main result is the confirmation of the remarkable quality
of the heavy-top limit as defined in \eqn{eq::heavytop} which agrees
with the $1/\mtop$ expansion to better than $0.5\%$ in the
phenomenologically interesting mass range $100\,{\rm GeV}\leq
\mhiggs\leq 300\,{\rm GeV}$, both at the \lhc{} as well as at the
Tevatron. This is an extremely comforting result because it validates
the numerous higher order analyses that have been carried out up to now,
in preparation for the \lhc{}
experiments~\cite{Ball:2007zza,Aad:2009wy}, and in particular for the
ongoing Higgs searches at the Tevatron~\cite{cdfd0:2009pt}.

It remains to be seen to what extent this finding carries over to less
inclusive quantities like distributions or phase space cuts. This
requires a substantial extension of our approach and is left for future
studies.

\paragraph{Acknowledgments.}
We would like to thank M.~Czakon, V.~del~Duca, S.~Forte, and
M.~Steinhauser for fruitful discussions. {\abbrev RVH} thanks the
organizers of the {\abbrev CERN TH} institute ``{\abbrev SM-BSM} physics
at the \lhc{}'', where part of this work was carried out, for
hospitality and financial support. This work was supported by {\abbrev
  DFG} under contract HA~2990/3-1, and the Helmholtz Alliance ``Physics
at the Terascale''.

\def\app#1#2#3{{\it Act.~Phys.~Pol.~}\jref{\bf B #1}{#2}{#3}}
\def\apa#1#2#3{{\it Act.~Phys.~Austr.~}\jref{\bf#1}{#2}{#3}}
\def\annphys#1#2#3{{\it Ann.~Phys.~}\jref{\bf #1}{#2}{#3}}
\def\cmp#1#2#3{{\it Comm.~Math.~Phys.~}\jref{\bf #1}{#2}{#3}}
\def\cpc#1#2#3{{\it Comp.~Phys.~Commun.~}\jref{\bf #1}{#2}{#3}}
\def\epjc#1#2#3{{\it Eur.\ Phys.\ J.\ }\jref{\bf C #1}{#2}{#3}}
\def\fortp#1#2#3{{\it Fortschr.~Phys.~}\jref{\bf#1}{#2}{#3}}
\def\ijmpc#1#2#3{{\it Int.~J.~Mod.~Phys.~}\jref{\bf C #1}{#2}{#3}}
\def\ijmpa#1#2#3{{\it Int.~J.~Mod.~Phys.~}\jref{\bf A #1}{#2}{#3}}
\def\jcp#1#2#3{{\it J.~Comp.~Phys.~}\jref{\bf #1}{#2}{#3}}
\def\jetp#1#2#3{{\it JETP~Lett.~}\jref{\bf #1}{#2}{#3}}
\def\jphysg#1#2#3{{\small\it J.~Phys.~G~}\jref{\bf #1}{#2}{#3}}
\def\jhep#1#2#3{{\small\it JHEP~}\jref{\bf #1}{#2}{#3}}
\def\mpl#1#2#3{{\it Mod.~Phys.~Lett.~}\jref{\bf A #1}{#2}{#3}}
\def\nima#1#2#3{{\it Nucl.~Inst.~Meth.~}\jref{\bf A #1}{#2}{#3}}
\def\npb#1#2#3{{\it Nucl.~Phys.~}\jref{\bf B #1}{#2}{#3}}
\def\nca#1#2#3{{\it Nuovo~Cim.~}\jref{\bf #1A}{#2}{#3}}
\def\plb#1#2#3{{\it Phys.~Lett.~}\jref{\bf B #1}{#2}{#3}}
\def\prc#1#2#3{{\it Phys.~Reports }\jref{\bf #1}{#2}{#3}}
\def\prd#1#2#3{{\it Phys.~Rev.~}\jref{\bf D #1}{#2}{#3}}
\def\pR#1#2#3{{\it Phys.~Rev.~}\jref{\bf #1}{#2}{#3}}
\def\prl#1#2#3{{\it Phys.~Rev.~Lett.~}\jref{\bf #1}{#2}{#3}}
\def\pr#1#2#3{{\it Phys.~Reports }\jref{\bf #1}{#2}{#3}}
\def\ptp#1#2#3{{\it Prog.~Theor.~Phys.~}\jref{\bf #1}{#2}{#3}}
\def\ppnp#1#2#3{{\it Prog.~Part.~Nucl.~Phys.~}\jref{\bf #1}{#2}{#3}}
\def\rmp#1#2#3{{\it Rev.~Mod.~Phys.~}\jref{\bf #1}{#2}{#3}}
\def\sovnp#1#2#3{{\it Sov.~J.~Nucl.~Phys.~}\jref{\bf #1}{#2}{#3}}
\def\sovus#1#2#3{{\it Sov.~Phys.~Usp.~}\jref{\bf #1}{#2}{#3}}
\def\tmf#1#2#3{{\it Teor.~Mat.~Fiz.~}\jref{\bf #1}{#2}{#3}}
\def\tmp#1#2#3{{\it Theor.~Math.~Phys.~}\jref{\bf #1}{#2}{#3}}
\def\yadfiz#1#2#3{{\it Yad.~Fiz.~}\jref{\bf #1}{#2}{#3}}
\def\zpc#1#2#3{{\it Z.~Phys.~}\jref{\bf C #1}{#2}{#3}}
\def\ibid#1#2#3{{ibid.~}\jref{\bf #1}{#2}{#3}}
\def\otherjournal#1#2#3#4{{\it #1}\jref{\bf #2}{#3}{#4}}
\newcommand{\jref}[3]{{\bf #1} (#2) #3}
\newcommand{\bibentry}[4]{#1, {\it #2}, #3\ifthenelse{\equal{#4}{}}{}{, }#4.}
\newcommand{\hepph}[1]{{\tt hep-ph/#1}}
\newcommand{\mathph}[1]{[math-ph/#1]}
\newcommand{\arxiv}[2]{{\tt arXiv:#1}}

\end{document}